\documentclass[final,3p,times]{elsarticle}
\usepackage[T1]{fontenc}
\usepackage[utf8]{inputenc}
\usepackage{amsmath}
\usepackage{amsfonts}
\usepackage{mathabx}
\usepackage{mathrsfs}
\usepackage{mathtools}
\usepackage{stmaryrd} 
\usepackage{hyperref}
\usepackage{scalerel,stackengine}
\usepackage{multirow}
\usepackage{relsize}
\usepackage{tocloft} 

\biboptions{sort&compress}

\stackMath
\newcommand\reallywidehat[1]{%
\savestack{\tmpbox}{\stretchto{%
  \scaleto{%
    \scalerel*[\widthof{\ensuremath{#1}}]{\kern-.6pt\bigwedge\kern-.6pt}%
    {\rule[-\textheight/2]{1ex}{\textheight}}
  }{\textheight}%
}{0.5ex}}%
\stackon[1pt]{#1}{\tmpbox}%
}
\usepackage{graphicx}
\usepackage[symbol]{footmisc}
\usepackage{tikz}
\newcommand*\circled[1]{\tikz[baseline=(char.base)]{
            \node[shape=circle,draw,inner sep=1.5pt] (char) {#1};}}

\newcommand{\convalgnospace}{\textit{convolution algorithm}}
\newcommand{\fftalgnospace}{\textit{fft algorithm}}

\newcommand{\convalg}{\textit{convolution algorithm} }
\newcommand{\fftalg}{\textit{fft algorithm} }

\begin{document}

\title{Hybrid-PFC: coupling the phase-field crystal model and its amplitude-equation formulation}

\author[TUD]{Maik Punke \footnote{Corresponding author: e-mail 
\href{mailto:maik.punke@tu-dresden.de}{maik.punke@tu-dresden.de}, 
     phone +49\,351\,463-41202,
     fax +49\,351\,463-37096}}
\author[TUD,DCMS]{Marco Salvalaglio}
\address[TUD]{Institute of Scientific Computing, TU Dresden, 01062 Dresden, Germany}
\address[DCMS]{Dresden Center for Computational Materials Science (DCMS), TU Dresden, 01062 Dresden, Germany}

\begin{abstract}
The phase-field crystal (PFC) model describes crystal structures at diffusive timescales through a periodic, microscopic density field. It has been proposed to model elasticity in crystal growth and encodes most of the phe\-no\-me\-no\-logy related to the mechanical properties of crystals like dislocation nucleation and motion, grain boundaries, and elastic or interface-energy anisotropies. To overcome limitations to small systems, a coarse-grained formulation focusing on slowly varying complex amplitudes of the microscopic density field has been devised. This amplitude-PFC (APFC) model describes well elasticity and dislocations while approximating microscopic features and being limited in describing large-angle grain boundaries. We present here seminal concepts for a hybrid multiscale PFC-APFC framework 
that combines the coarse-grained description of the APFC model in bulk-like crystallites while exploiting PFC resolution at dislocations, grain boundaries, and interfaces or surfaces. This is achieved by coupling the two mo\-dels via an advanced discretization based on the Fourier spectral method and allowing for local solution updates. This discretization also generalizes the description of boundary conditions for PFC models. We showcase the framework capabilities through two-dimensional benchmark simulations. We also show that the proposed formulation allows for overcoming the limitations of the APFC model in describing large-angle grain boundaries.
\end{abstract}

\addcontentsline{toc}{chapter}{Hybrid-PFC: coupling the phase-field crystal model and its amplitude-equation formulation}
\maketitle   

\section{Introduction}

Crystalline materials are ubiquitous in nature and technology. To describe them comprehensively, one must consider several key phenomena, including solidification, capillarity, elastic and plastic deformations, the formation of grain boundaries (GBs), and anisotropies related to crystallographic directions. These phenomena and features typically require descriptions at different time and length scales, presenting challenges in selecting appropriate models and methods for comprehensive investigations \cite{rollett2015understanding}. 

While lattice-dependent features such as anisotropies and defect structures \cite{hoyt2003atomistic} can be evaluated by microscopic approaches, the growth of crystals or evaluation of mechanical properties involves long time scales and large length scales, typically not accessible within these methods. On the other hand, continuum approaches proved successful in describing macroscopic behaviors, for instance, via advanced front-tracking or phase-field methods for crystal growth \cite{KOBAYASHI1993410,Karma1998,Zhu2007,Steinbach2009,Pan2010,takaki2014phase,kaiser2020} coping with large systems and long timescales. However, they usually lack a direct connection to the lattice symmetry and microscopic features. Lattice-dependent effects can then be partially described and included mainly through parameters and additional functions, e.g., anisotropic interface energies \cite{SUZUKI2002125,Tor2009}.

The so-called phase-field crystal (PFC) model \cite{Elder2002,Elder2004,Provatas2010,Emmerich2012}
emerged as a prominent approach to describe crystal structures at large (diffusive) timescales through a continuous, periodic order parameter representing the atomic density. Although representing a minimal, phenomenological approach, it reproduces the main phenomenology for crystalline systems, from crystal growth to lattice deformation and dislocation kinematics, and it is suitable for both two and three-dimensional investigation \cite{
Emmerich2012}. Also, it allows for a self-consistent description of anisotropies resulting from the lattice structure \cite{PODMANICZKY2014148,Ofori-Opoku2018}, and has been developed towards advanced modeling of elastic relaxation \cite{stefanovic2006phase,toth2013nonlinear,heinonen2016consistent,skaugen2018separation,skogvoll2022hydrodynamic}.

 The amplitude expansion of the PFC model (APFC) was developed to overcome the length scale limitations in the PFC framework~\cite{goldenfeld2005renormalization,athreya2006renormalization,salvalaglio2022coarse}. In particular, the continuous density in PFC models is replaced by the (complex) amplitudes of sets of Fourier modes (wave vectors), reproducing a targeted crystal symmetry. This coarse-grained version of the PFC has been used to examine many phenomena, including liquid/solid fronts, dislocation nucleation and motion, as well as strained films. We refer to Ref.~\cite{salvalaglio2022coarse} for a recent review. However, the APFC is limited to small rotation angles with respect to a reference lattice, which prevents an accurate description of large-angle GBs \cite{salvalaglio2022coarse,huter2017modelling,spatschek2010amplitude}. Although requiring sophisticated numerical implementation and so far being demonstrated for very simple systems, some approaches were developed to overcome this issue~\cite{athreya2007adaptive,bervcivc2018adaptive,bervcivc2020enabling}. It remains, however, that the APFC model proves too coarse to inspect accurately microscopic effects while reaching large scales.

This paper proposes a hybrid multiscale PFC-APFC framework 
with PFC accuracy in regions of interest (e.g., at defects and interfaces) while exploiting the coarse resolution of the APFC model elsewhere. 
In the proposed method, these two models are consistently coupled, leveraging an advanced pseudo-spectral method for spatial discretization. We showcase the capabilities of this newly proposed method via selected numerical investigations that focus on GB structures and the anisotropic solidification of two-dimensional crystals. Benchmarks against standard PFC and APFC models are reported. We also show a proof of concept for its application to describe large-angle GBs, providing a practical solution to a crucial limitation of the APFC model.

The manuscript is organized as follows. In section~\ref{sec:pfc}, we introduce the basics of the PFC model. We also present a method to numerically solve the corresponding evolution equation within the pseudo-spectral framework by building on real-space convolutions, which is convenient for the targeted coupling with the APFC model and is also shown to allow for various boundary conditions beyond the classical periodic one considered by usual Fourier pseudo-spectral implementations. Section~\ref{sec:apfc} introduces the APFC model. We then present our hybrid framework in section~\ref{sec:hybrid}. We show that it allows for PFC accuracy while generally using coarser resolution with respect to standard approaches. It is also demonstrated that this hybrid-PFC approach allows for overcoming the limit for large-angle GBs in the APFC model for bicrystals but still allows for a faster computation than the classical PFC. Finally, we draw our conclusions in section~\ref{sec:conclusions}. Technical details about the implementation, auxiliary convergence studies, and further additional information are reported in~\ref{sec:appendixTable} and ~\ref{sec:appendix}. 

\section{Phase-field crystal modeling}
\label{sec:pfc}
The phase-field crystal (PFC) model \cite{Elder2002,Elder2004,Emmerich2012}
describes crystal structures at diffusive timescales through a continuous, periodic order parameter $\psi \equiv \psi(\mathbf{r},t)$ representing the atomic density. It is based on a Swift-Hohenberg-like free energy functional \cite{Elder2002,Elder2004,Emmerich2012}
    \begin{equation}
F\left[\psi\right] = \int_\Omega \left( \dfrac{\lambda-\kappa}{2}\psi^2 -  \delta \dfrac{\psi^3}{6} + \dfrac{\psi^4}{12}+ \dfrac{\kappa}{2}\psi\, \mathcal{L}\, \psi\right) \,\, \rm{d}\mathbf{r},   
    \label{eq:F1}
    \end{equation}
with $\delta,\,\lambda,\,\kappa \geq 0$ parameters characterizing the phase space and material properties together with the global average density $\Psi_0=\frac{1}{|\Omega|}\int_\Omega \psi  \rm{d}\mathbf{r}$, and $\Omega \in \mathbb{R}^n$ the domain of definition of $\psi$ with $n=2$ defined here as $\Omega =[-L_x/2,L_x/2]\times [-L_y/2,L_y/2]$. $\mathcal{L}=\prod_{i=1}^Q (q_i^2 + \nabla^2)^2$ is a differential operator that approximates a two-point correlation function and thus encodes the crystal symmetry, with $q_i$ characteristic wavenumbers \cite{Emmerich2012}. 2D triangular symmetry can be modeled with $\mathcal{L}=(1+\nabla^2)^2$ and 2D square symmetry with $\mathcal{L}=(1+\nabla^2)^2(2+\nabla^2)^2$. Table ~\ref{tab:numerics} summarizes the model parameters for all simulations in this paper. 

The dynamical equation for $\psi$ is described via a conservative ($\mathrm{H}^{-1}$) gradient flow of $F$,
    \begin{equation}
    \begin{split}
\partial _t \psi &= M\nabla^2 \dfrac{\delta F\left[\psi\right]}{\delta \psi}=L[\psi]+ N[\psi],
    \end{split}
    \label{eq:pfc}
    \end{equation}
with $L$ and $N$, the linear and non-linear terms reading
    \begin{equation}
    \begin{split}
    L[\psi] &= M(\lambda-\kappa+\kappa\mathcal{L})\nabla^2 \psi,\\
    N[\psi]&= M\nabla^2(-\dfrac{\delta}{2} \psi^2 + \dfrac{\psi^3}{3}).
       \label{eq:pfc2}
       \end{split}
    \end{equation}
  The parameter $M>0$ corresponds to a mobility factor, which is assumed to be constant. Further extensions of~\eqref{eq:F1}-\eqref{eq:pfc} may be readily considered to account for other lattice symmetries \cite{Greenwood2010,Greenwood2011,Mkhonta2013} (in both 2D and 3D). 

\subsection{Fourier pseudo-spectral method}

The partial differential Eq.~\eqref{eq:pfc} can be solved with different methods. Given the high differential order (6th or higher in PFC models) and the need to resolve the microscopic density everywhere, which is easily addressed by uniform spatial discretizations, a widely adopted and convenient approach consists of using a Fourier pseudo-spectral method \cite{cheng2008efficient,cheng2019energy,wise2009energy,gomez2012unconditionally,Pinomaa_2024,skogvoll2024comfit}. This approach usually enforces periodic boundary conditions (PBC) and allows for stable integration of the equations with larger timesteps than real-space counterparts.

We denote by $k$ the $L_2$-norm of the spatial Fourier vector $\mathbf{k}=(k_1,k_2)$, and $\widehat{f}$ is the spatial Fourier transform of the periodic function $f$. 
With this notation, we may write the equation for $\partial_t \hat{\psi}$ from \eqref{eq:pfc} as
\begin{equation}\label{eq:pfc_k}
\partial _t \widehat{\psi}= \widehat{L[\psi]} + \widehat{ N[\psi]},
\end{equation}
with
    \begin{equation}
    \begin{split}
\widehat{L[\psi]} &=  K \widehat{\psi},
   \\
\widehat{ N[\psi]} &= -Mk^2 \left(-\dfrac{\delta}{2}\widehat{\psi^2}+\dfrac{\widehat{\psi^3}}{3} \right),
    \end{split}
    \label{eq:operator2}
    \end{equation}
and $K$ a polynomial in $k$. For the energy functional encoding  triangular crystal symmetry, it reads:
    \begin{equation}
K(k)=  M(-\lambda k^2+ 2\kappa k^4 - \kappa k^6).
    \end{equation}
    For the energy functional describing square crystal symmetry, $K$ takes the form
    \begin{equation}
K(k)=  M(-(\lambda+3\kappa) k^2+ 12\kappa k^4 -13 \kappa k^6+6\kappa k^8 -\kappa k^{10}).
    \end{equation}
Eq.~\eqref{eq:pfc_k} can then be integrated to compute $\widehat\psi(t)$. A simple but effective approach consists of the semi-implicit scheme 
    \begin{equation}
\prescript{s+1}{}{\widehat{\psi}} =   \dfrac{\prescript{s}{}{\widehat{\psi}} + \Delta t  \widehat{N[\prescript{s}{}{\psi}]}}{1-\Delta t K}\quad \fftalg
    \label{eq:fft}
    \end{equation}
with $\Delta t$ the (constant) time step size and $s$ labelling the time step such that $t=s\Delta t$. Note that this scheme requires one global Fourier- and one inverse Fourier transformation after each time step update, as $\prescript{s}{}{\psi}$ must be evaluated to compute $N[\prescript{s}{}{\psi}]$. The (inverse) Fourier transformations can be efficiently computed using the FFTW library \cite{frigo1998fftw}.
This standard approach is hereafter referred to as the \fftalgnospace.

\subsection{Local updates by the convolution algorithm}
Performing an additional inverse Fourier-transform on the integration scheme \eqref{eq:fft} and using the convolution theorem~\cite{katznelson2004introduction}, we can alternatively write it in real space
\begin{equation}
\begin{split}
\prescript{s+1}{}{\psi} &=   \left[\prescript{s}{}{\psi} +  (-\dfrac{\delta}{2} \prescript{s}{}{\psi^2} + \dfrac{\prescript{s}{}{\psi^3}}{3})\raisebox{.5pt}{\textcircled{\raisebox{-2.5pt} {*}}} Q \right]\raisebox{.5pt}{\textcircled{\raisebox{-2.5pt} {*}}}R \quad \convalg,\\
Q&=\Delta t M \widehat{k^2},\\
R&=-\widehat{\dfrac{1}{1-\Delta t K}}
       \end{split}
    \label{eq:conv}
    \end{equation}
with $\raisebox{.5pt}{\textcircled{\raisebox{-2.5pt} {*}}}$ being the cyclic convolution. See~\ref{sec:linconv} and~\ref{sec:cycconv} for its detailed definition and implementation. $Q$ and $R$ represent discretized operators and do not change during a simulation. In this approach, two (cyclic) convolutions need to be evaluated in each step instead of two Fourier transformations per time step. Both the schemes \eqref{eq:fft} plus an inverse Fourier transform and \eqref{eq:conv} are equivalent in that they provide an integration scheme for \eqref{eq:pfc} delivering $^{s+1}\psi$. 
It should be emphasized that the equivalence between the \fftalg and the \convalg is not restricted to the IMEX time-stepping scheme used in~\eqref{eq:fft} and also holds for additional numerical stabilization techniques as in Refs.~\cite{elsey2013simple,punke2023improved}.

Importantly, computing convolutions is typically expensive, especially compared with methods involving Fourier transforms computed via the Fast Fourier Transform algorithm. However, two scenarios are relevant for the present discussion and the scope of this work, where the \convalg has advantages compared to the \fftalgnospace:
\begin{itemize}
    \item \textit{Local updates of the density field}. Contrary to the global \fftalgnospace, convolutions can be computed locally, allowing for a local update of the density field. More technical details are reported in~\ref{sec:linconv}.
    \item \textit{Boundary conditions other than periodic}. By replacing the cyclic convolutions $\raisebox{.5pt}{\textcircled{\raisebox{-2.5pt} {*}}}$  by linear convolutions $*$, other types of boundary conditions can be used, e.g., Dirichlet BC (DBC), see a specific discussion in ~\ref{sec:linconv} and~\ref{sec:cycconv}. 
\end{itemize}
The following section provides examples that further elaborate on these properties.

 \subsection{Examples: fft vs convolution algorithms}

\begin{figure}[h!]
\centering
\includegraphics[width=\textwidth]{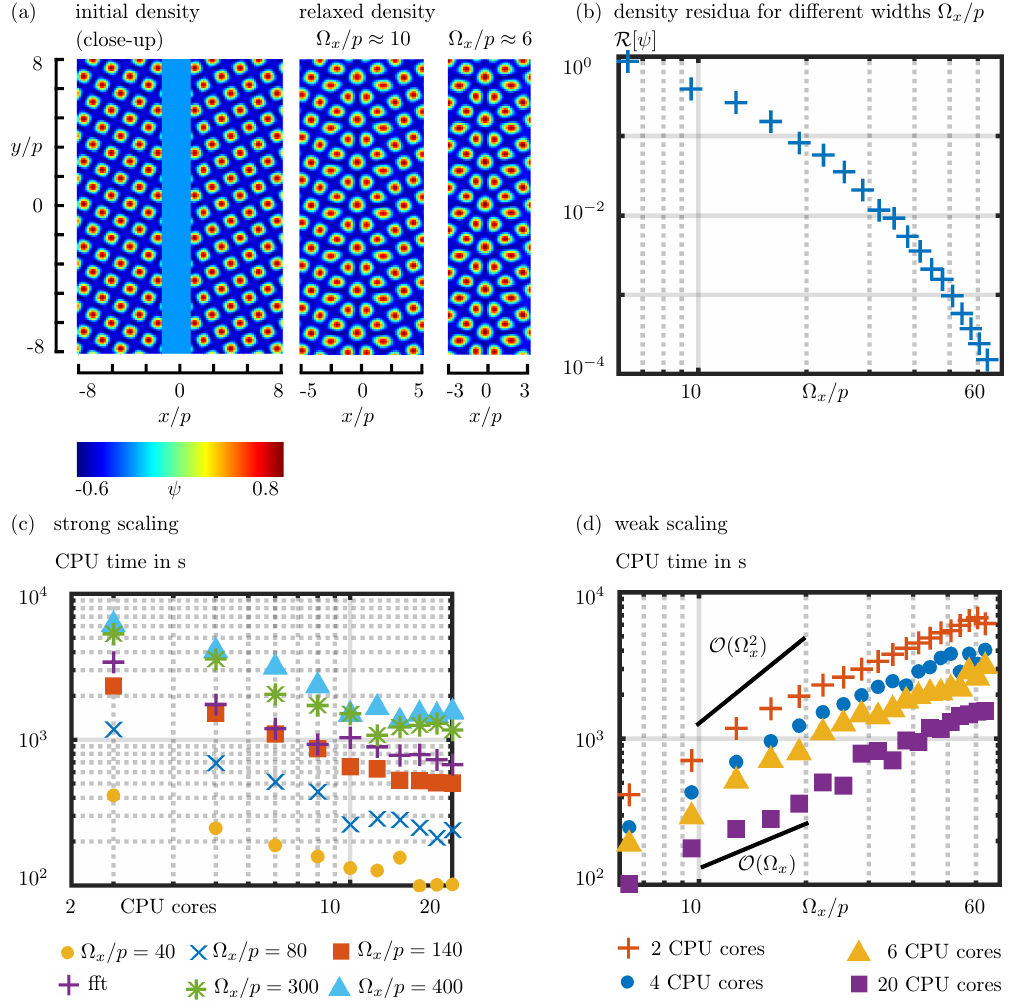}
\caption{Comparison of performances between the \convalg~\eqref{eq:conv} and \fftalg~\eqref{eq:fft}. The formation of a symmetric $\Sigma 5$ GB with underlying square symmetry is considered. (a) Magnification of the initial and relaxed structures for different widths $\Omega_x$, obtained with the \convalgnospace. (b) Numerical residua of the \convalg for different widths $\Omega_x$, computed as the squared $L_2$ distance to the fft solution in the central region. (c) Strong scaling of the \convalg and the \fftalgnospace. (d) Weak scaling of the \convalg where black lines illustrate linear and quadratic reference slopes. Lengths are scaled with the atomic spacing $p$.}
	\label{fig:GB}
    \end{figure}

As the first example, we consider a setting where only one part of the domain needs to be updated and show that the \convalg can outperform the \fftalgnospace; see Fig.~\ref{fig:GB}. We simulate the formation of a symmetric $\Sigma 5$ GB in a bicrystal with underlying square symmetry ($Q=2$, $q_1=1$ and $q_2=\sqrt{2}$ entering $\mathcal{L}$ in Eq.~\eqref{eq:F1}). The initial condition consists of two bulk crystals with a relative (symmetric) tilt angle of $\pm 26.55^\circ/2 $ and a liquid stripe in between; see Fig.~\ref{fig:GB}(a). $\Omega$ is set to a multiple of the period of $\psi$ along $x$- and $y$-directions as imposed by the rotated lattice vectors to model a strain-free system~\cite{mellenthin2008phase,blixt2021evaluation} (we set $L_x/2 = 4215$ and $L_y/2 = 49$, imposing  $\Omega \approx 1342\times 16$ UC with 1 UC being the smallest repeat unit of $\psi$: 1UC $= [0,p]\times [0,p]$ with $p=2\pi$).  Model and numerical parameters are chosen as detailed in Table~\ref{tab:numerics}. 

As expected, after an initial relaxation of the initial condition ($t=10^3$), a $\Sigma 5$ GB forms. In Fig.~\ref{fig:GB}(b)-(d), we benchmark the \convalg against the \fftalg under PBC. For the former we vary the size of the region where the density field is updated $[-\Omega_x/2,\Omega_x/2]\times [-L_y/2,L_y/2]$ within the range $6.3p\leq \Omega_x\leq 63p$ and measure the resulting density residual $\mathcal{R}[\psi]$ with respect to the \fftalg as the squared $L_2$ distance to the fft solution in the region around the formed GB $[-1.6p,1.6p]\times [-L_y/2,L_y/2]$. By increasing the extension of $\Omega_x$ the residuum $\mathcal{R}[\psi]$ decreases exponentially; see Fig.~\ref{fig:GB}(b). Furthermore, we analyze the weak and strong scaling of the \convalg to benchmark the computational costs against the \fftalgnospace, evaluated as the required CPU time for $t=10^3$, see Fig.~\ref{fig:GB}(c)-(d).   Independent of the number of CPU cores, the \convalg performs better than the \fftalg for $\Omega_x\leq 22.2p$. A saturation of the computation costs is observed for $\approx 10$ CPU cores for both the \convalg and the \fftalgnospace, see Fig.~\ref{fig:GB}(c). In the weak scaling analysis, the \convalg shows a linear to quadratic increase in the computation cost for increasing $\Omega_x$, see Fig.~\ref{fig:GB}(d). We point out that the weak scaling analysis was not conducted for the \fftalg as it requires an update of the whole computational grid at each time step.  Further run-time improvements can be achieved for the \convalg if a singular value decomposition (SVD) is used. In Fig.~\ref{fig:svd}(a) in the Appendix, we present a detailed study of the SVD. If not further specified, all numerical simulations are conducted with $20$ CPU cores in the following.
 
\begin{figure}[h!]
\centering
\includegraphics[width=\textwidth]{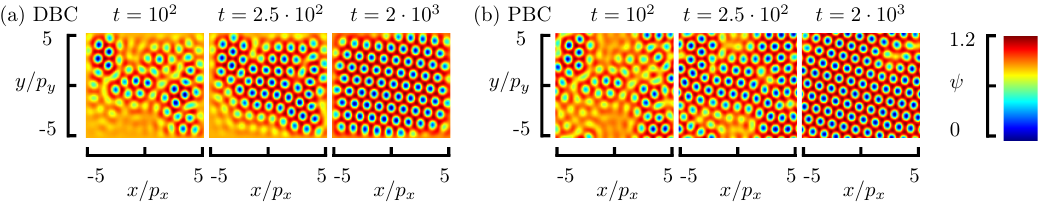}
\caption{Examples of solidifications with different boundary conditions. (a) DBC ($\left. \psi \right| _{\partial \Omega}\equiv \Psi_0$) obtained through the \convalg~\eqref{eq:conv} with linear instead of cyclic convolutions and (b) PBC obtained by the \convalg~\eqref{eq:conv} or the \fftalg~\eqref{eq:fft}. The initial conditions were chosen in both panels as the same random field. Lengths are scaled with the atomic spacings along the $x$- and $y$-axis, $p_x$ and $p_y$.}
\label{fig:random}
\end{figure}

As a second example, we showcase the ability of the convolution algorithm to model other types of boundary conditions, e.g., Dirichlet BC (DBC), such that $\left. \psi \right| _{\partial \Omega}\equiv \Psi_0$, corresponding to constraint the density to its average values at the boundary. Technically, this requires linear convolutions instead of cyclic convolutions or a modification of the \fftalgnospace; see the Appendix for further implementation details~\ref{sec:linconv}. As a simple setup, we initialize a random field and let the system relax with a PFC model reproducing triangular symmetry ($Q=1$ and $q_1=1$ entering $\mathcal{L}$ in Eq.~\eqref{eq:F1}), see parameters in Table~\ref{tab:numerics}. The domain is set to $L_x/2=18.13$ and $L_y/2=15.7$, imposing $\Omega = 10\times 10$UC with 1UC $= [0,p_x]\times [0,p_y]$ and $p_x=2/\sqrt{3} p_y= 4\pi/\sqrt{3}$. In Fig.~\ref{fig:random}, we compare the solidification process under different boundary conditions.  While with DBC, the numerical simulation mimics growth under confinement with the growth front stopping at the boundary, see.~Fig.~\ref{fig:random}(a), a solidification across the boundary is visible with PBC, see Fig.~\ref{fig:random}(b). For the modeling of PBC, cyclic convolutions or the \fftalg are necessary, further discussions are reported in ~\ref{sec:cycconv}.

\section{Amplitude phase-field crystal modeling}
\label{sec:apfc}
The microscopic density $\psi$, described by PFC model~\eqref{eq:pfc}, is a smooth, periodic field. Therefore, it is well described by its principal Fourier components~\cite{goldenfeld2005renormalization,athreya2006renormalization,salvalaglio2022coarse},
\begin{equation}
\psi\approx \psi_0+\sum_{m=1}^N \left( \eta_m \mathrm{e}^{i\boldsymbol{q}_m\cdot \boldsymbol{r}} + \mathrm{c.c.} \right),
    \label{eq:1mode}
\end{equation}
with imaginary unit $i$, local average density $\psi_0$, complex amplitudes $\{\eta_m\}_{m=1}^N$ and reciprocal lattice vectors $\{\boldsymbol{q}_m\}_{m=1}^N$ reproducing different crystal symmetries in two or three dimensions~\cite{salvalaglio2022coarse}. c.c. denotes the complex conjugate. 
For instance, a crystal with triangular symmetry in 2D can be described by setting $\{\boldsymbol{q}_m\}_{m=1}^3=q_0\{(0,1),(\pm\sqrt{3}/2,-1/2)\}$. This corresponds to considering the shortest reciprocal space vector only in a 1-Mode formulation ("mode" refers to a family of equal-length $\boldsymbol{q}$). Square symmetric crystals can be described by a 2-Mode approximation with reciprocal space vectors $\{\boldsymbol{q}_m\}_{m=1}^4=q_0\{(0,1),(1 ,0),(1 ,\pm1)\}$ \cite{salvalaglio2022coarse}. For simplicity, we consider hereafter crystals described only by a triangular 1-Mode approximation.

Through a renormalization group approach~\cite{goldenfeld2005renormalization}, or equivalently, by substituting the approximation \eqref{eq:1mode} into \eqref{eq:F1} and integrating over the unit cell, it is possible to derive the free energy for complex amplitudes $\{\eta_m\}_{m=1}^3$ with constant $\psi_0$~\cite{athreya2006renormalization,salvalaglio2022coarse}
\begin{equation}
    \mathcal{F}[\psi_0,\eta_k]=\int_\Omega \left[ \dfrac{\lambda-\kappa-\delta \psi_0 +\psi_0^2}{2}\Phi+\dfrac{1}{4}\Phi^2+\sum_{m=1}^3  \left(\kappa  \left(G_m \eta_m\right)^2 -\dfrac{1}{2}|\eta_m|^4\right) +f_s+E\right]\, \rm{d}\mathbf{r},
\end{equation}
with 
\begin{equation}
    \begin{split}
    \Phi&=2\sum_{m=1}^3 \eta_m \eta_m^*,\\
    G_m &= \nabla^2+2i \boldsymbol{q}^m\cdot \nabla\quad \forall m=1,\dots, 3,\\
        f_s&= (-\delta+2\psi_0)\left( \prod_{m=1}^3 \eta_m+ \prod_{m=1}^3 \eta_m^*\right),\\
E&= \lambda\dfrac{\psi_0^2}{2}-\delta\dfrac{\psi_0^3}{6}+\dfrac{\psi_0^4}{36}.
\end{split}
\label{eq:apfc_energy}
\end{equation}
$\Phi$ corresponds to an order parameter constant in bulk phases while decreasing at defects and interfaces. In Ref.~\cite{yeon2010density}, this framework has then been extended to $\psi_0$ fields varying similarly to complex amplitudes, so still slowly varying compared to the underlying microscopic density field $\psi$, and is based on the same free energy $F$ under assumptions which allow for neglecting gradients of $\psi_0$.

The evolution of $\psi$ as in Eq.~\eqref{eq:pfc} is then approximated through the evolution of $\{\eta_m\}_{m=1}^N$ dictated by a weighted non-conservative ($L^2$) gradient flow of $\mathcal{F}$ coupled to the evolution of 
$\psi_0$ dictated by a conservative ($H^{-1}$) gradient flow of $\mathcal{F}$ \cite{salvalaglio2022coarse,yeon2010density}
\begin{equation}
\begin{split}
    \partial_t \eta_m  &= -|q_m|\dfrac{\delta \mathcal{F}}{\delta \eta_m^*} \\
    &=-\left(\kappa G_m^2+\lambda-\kappa -\delta \psi_0 + \psi_0^2\right)\eta_m-\left((\Phi-|\eta_m|^2)\eta_m +(-\delta+2\psi_0)\prod_{j\neq m}^3 \eta_j^*\right)\quad \forall m=1,\dots, 3,\\
    \partial_t \psi_0 &=\nabla^2 \dfrac{\delta \mathcal{F}}{\delta \psi_0}\\
    &=\nabla^2\left[(-\delta +2\psi_0)\dfrac{\Phi}{2}+2 \left(\prod_{m=1}^3 \eta_m + \prod_{m=1}^3\eta_m^*\right)+\lambda \psi_0-\psi_0^2\dfrac{\delta}{2}+\dfrac{\psi_0^3}{3}\right].
    \end{split}
    \label{eq:apfc}
\end{equation}

\subsection{Numerical simulations}

The APFC model is efficiently solved by real space methods, which allow for exploiting adaptive refinements \cite{bervcivc2018adaptive,bervcivc2020enabling,athreya2007adaptive,Salvalaglio2017,Praetorius2019}. Indeed, these approaches allow for fine spatial discretizations at defects and interfaces while using coarse discretizations elsewhere. In this work, however, we target the coupling of the APFC model with solutions from the PFC model in the region where fine spatial discretization would be needed. In the domain where we aim to solve the APFC within the hybrid-PFC framework, discussed in Sect.~\ref{sec:hybrid}, we can then use a coarse uniform spatial discretization, analogous to the coarsest considered in adaptive methods. This can be handled efficiently by a Fourier spectral method as considered for the PFC model in Sect.~\ref{sec:pfc}.

In particular, numerical solutions of Eqs.~\eqref{eq:apfc} can be obtained by using a Fourier pseudo-spectral discretization in space combined with a linear first-order IMEX time-stepping scheme, similar to~\eqref{eq:pfc_k}, namely
\begin{equation}
\begin{split}
\frac{\prescript{s+1}{}{\widehat{\eta_{m}}}-\prescript{s}{}{\widehat{\eta_m}}}{\Delta t}&=\prescript{s+1}{}{\widehat{L_{\eta_m}}}+\prescript{s}{}{\widehat{N_{\eta_m}}}\quad \forall m=1,\dots, 3,\\
\frac{\prescript{s+1}{}{\widehat{\psi}_{0}}-\prescript{s}{}{\widehat{\psi}_{0}}}{\Delta t}&=\prescript{s+1}{}{\widehat{L_{\psi_0}}}+\prescript{s}{}{\widehat{N_{\psi_0}}}, 
\end{split}
\label{eq:apfc_fft}
\end{equation}
with, considering here a 1-mode approximation of the triangular lattice as mentioned above,
\begin{equation}
\begin{split}
\widehat{L_{\eta_m}}&=\left[ \kappa-\lambda -\kappa  (-k^2+2i\boldsymbol{q}^m\cdot \boldsymbol{k})^2\right] \widehat{\eta_m}\quad \forall m=1,\dots, 3,\\
\widehat{N_{\eta_m}}&=\delta \widehat{\psi_0\eta_m} -\widehat{ \psi_0^2\eta_m}-\widehat{\Phi\eta_m}+\widehat{\eta_m\eta_m^*} + \delta \reallywidehat{\prod_{j\neq m}^3 \eta_j^*} -2 \reallywidehat{\psi_0\prod_{j\neq m}^3 \eta_j^*} \quad \forall m=1,\dots, 3,\\
\widehat{L_{\psi_0}}&=-\lambda k^2 \widehat{\psi_0},\\
\widehat{N_{\psi_0}}&=k^2\left[\dfrac{\delta}{2}\widehat{\Phi} -\widehat{\psi_0\Phi}-2\left(\reallywidehat{\prod_{j=1}^3 \eta_j }+ \reallywidehat{\prod_{j=1}^3\eta_j^*}\right)+\dfrac{\delta}{2}\widehat{\psi_0^2}-\widehat{\dfrac{\psi_0^3}{3}}\right].
\end{split}
\end{equation}

\subsection{Example: growing crystal}

\begin{figure}[b!]
\centering
\includegraphics[width=\textwidth]{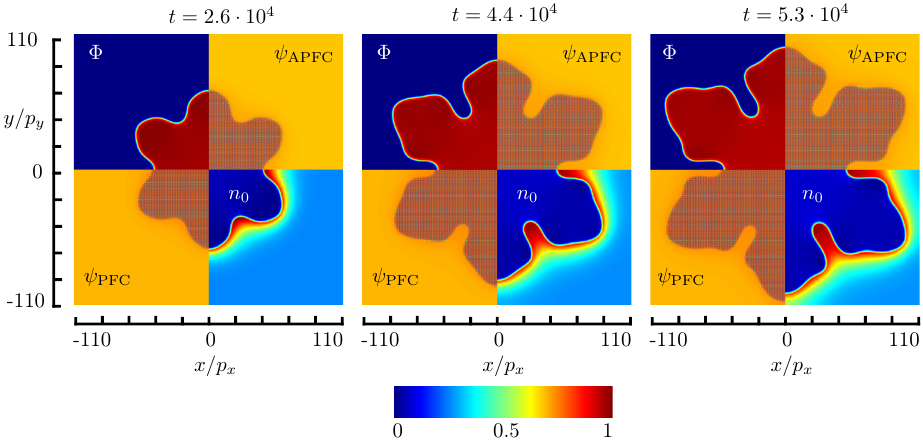}
\caption{APFC vs. PFC model: Dendritic solidification in an undercooled melt at different time steps. For the APFC model, the order parameter $\Phi$, local average density $\psi_0$ and reconstructed density $\psi_\text{APFC}$ are plotted and compared to the density obtained by the PFC model $\psi_\text{PFC}$. We interpolated all quantities on a fine grid for illustration purposes and normalized them from 0 to 1. The quantities $\psi_\text{PFC}$, $\Phi$, $\psi_0$ and $\psi_\text{APFC}$ vary in the following intervals: $0.03\leq \psi_\text{PFC}\leq 1.21$, $0\leq \Phi\leq 0.10$, $0.83 \leq \psi_0 \leq 0.89$ and $0.05\leq \psi_\text{APFC}\leq 1.23$. Lengths are scaled with the atomic spacings along the $x$- and $y$-axis, $p_x$ and $p_y$.}
\label{fig:pfc_apfc}
\end{figure}

Aiming at a coupling between PFC and APFC, we first consider a setting to benchmark how similar the description achieved by these models is. We look at a benchmark simulation of a growing triangular crystal seed in an undercooled melt \cite{punke2022explicit}. Model and simulation parameters are reported in Table~\ref{tab:numerics}. We simulate this process with the APFC model and compare it to the same setting explicitly obtained by integrating the equation of the PFC model with $M=0.66$ (the mobility $M$ was chosen to have compatible time scales of the APFC and the PFC model). In Fig.~\ref{fig:pfc_apfc}, different time steps of the solidification process are plotted with snapshots of both PFC ($\psi_\text{PFC}$) and APFC ($\psi_\text{APFC}$, $\psi_0$, and $\Phi$) solutions. Therein, $\psi_\text{APFC}$ is obtained by reconstructing the density field from $\psi_0$ and $\{\eta_m\}_{m=1}^3$ via Eq.~\eqref{eq:1mode}. Both models lead to dendritic solidification, with very similar morphologies for the solid-liquid interface. We note a minor difference in the approaches: the PFC model produces a slightly enlarged lattice spacing during growth, which is negligible in the APFC simulation. This results in a (small) positive hydrostatic mechanical stress inside the solid phase for the used parameter combination; see Refs.~\cite{punke2022explicit,punke2023evaluation} for further quantifications and discussions. The main advantage of the APFC compared to the PFC model, owing to its coarse-grained nature, is its scaling property. While the PFC model needs a relatively fine mesh resolution, e.g. for the simulation in Fig.~\ref{fig:pfc_apfc} the maximal possible resolution (see also \ref{sec:refinement_domain}) of $\Delta x =1.33$, $\Delta y = 1.20$ is used leading to $\approx 1.3\cdot 10^4$s simulation time, the APFC can be modeled with a maximal possible resolution of $\Delta x = 8.90 $ and $\Delta y=7.70$, leading to $\approx 1.4\cdot 10^3$s simulation time, see Table~\ref{tab:numerics}. By comparing the APFC and PFC solutions with their respective refined numerical solutions for $\Delta x =0.50$ and $\Delta y = 0.43$, we make sure to reach convergent results (see the Appendix for a refinement and run-time study as well as for further explanations concerning the critical grid sizes~\ref{sec:refinement_domain}). For illustrating purposes, in particular, to show the results with the same resolution of the density peaks, we interpolated both the APFC as well as the PFC solution in Fig.~\ref{fig:pfc_apfc} by Fourier padding as post-processing (further explanation in Appendix~\ref{sec:interpol}).

 Overall, it is shown that for the setup illustrated in Fig.~\ref{fig:pfc_apfc}, the APFC leads to comparable results to the PFC model while allowing for coarser computational grids and, therefore, reduced computational costs ($\approx 10$ times faster run-time of the APFC compared to the PFC). For systems featuring small deviations from bulk settings and crystal growth, the coupling of these models leveraging the fields reported in Fig.~\ref{fig:pfc_apfc} can then be envisaged. Note that the comparison and the discussion of performances and required resolution are shown here for a system featuring an extended interface. For bulk systems in elastic regimes, the gain in the computational costs is expected to increase significantly: the ideal but instructive limit of a bulk, relaxed crystal would still require resolving the oscillation of the microscopic density when considering the PFC model, while amplitudes entering the APFC could be even determined analytically (ideally corresponding to one discretization point). However, the APFC is in quantitative agreement with the PFC only for the growth of (almost) relaxed crystals or in the presence of small deformations. Below, we will show how the hybrid-PFC method we propose in this work may overcome these limitations. We remark that the good matching of PFC and APFC results, particularly when considering solid-liquid (or phase in general) interfaces, can be achieved only when considering a spatially varying average density $\psi_0$. The widely adopted simplification of a constant $\psi_0$ is found to work well for bulk and solid systems (see, e.g., \cite{benoitmarechal2024mesoscale}), but nearly-isotropic solidification front would be obtained, with weaker growth anisotropy than the PFC counterpart, no dendritic shapes and significantly faster solidification dynamics.

\section{Hybrid-PFC Model}
\label{sec:hybrid}
\subsection{Hybrid simulation scheme}
We introduce here the \textit{hybrid-PFC} model and showcase its capabilities via selected numerical benchmarks against the original PFC 
and APFC models, namely Eqs.~\eqref{eq:pfc} and \eqref{eq:apfc} respectively.
Within the hybrid-PFC model, we define the PFC density $\psi_\text{PFC}$ (PFC part of the hybrid-PFC model), the APFC amplitudes $\{\eta_m\}_{m=1}^3$ and the local average density $\psi_0$  (APFC part of the hybrid-PFC model) on a (large) domain $\Omega$ with different numerical discretizations (fine grid for $\psi_\text{PFC}$, coarse grids for $\{\eta_m\}_{m=1}^3$ and $\psi_0$). Only in a (small) subdomain $\omega\subset \Omega$, $\psi_\text{PFC}$ is updated by the \convalg~\eqref{eq:conv} with linear convolutions, whereas $\{\eta_m\}_{m=1}^3$ and $\psi_0$ are updated everywhere in the domain $\Omega$ by~\eqref{eq:apfc_fft}. A consistent coupling between the PFC and APFC within the hybrid-PFC model is ensured by an interpolation on a subdomain $\widetilde{\omega}$ with $\omega\subset \widetilde{\omega} \subset\Omega$; see also Fig.~\ref{fig:hybrid_scetch}(b).

The algorithm can be summarized as follows. At every time step $s\geq 0$, we 
\begin{equation}
  \text{full }  \left\{
  \begin{array}{l}
  \text{simplified }  \left\{ \begin{array}{ll}
      \circled{1} \text{ compute } \left\{ \prescript{s+1}{}{\eta_m}\right\}_{m=1}^3 \text{ and } \prescript{s+1}{}{\psi_0} &\textit{ (global APFC update)}   \\
      \\
       \circled{2} \text{ compute }  \prescript{s+1}{}{\psi_\text{PFC}} \text{ on } \omega &\textit{ (local PFC update)}\\
       \\
        \circled{3}  \text{ interpolate } \left\{ \prescript{s+1}{}{\eta_m}\right\}_{m=1}^3, \prescript{s+1}{}{\psi_0}  \text{ on } \widetilde{\omega}  \text{ and compute } \prescript{s+1}{}{\psi_\text{APFC}} &\textit{ (local APFC interpolation)}\\
        \\
    \circled{4}   \text{ correct }\prescript{s+1}{}{\psi_\text{PFC}} \text{ by }\prescript{s+1}{}{\psi_\text{APFC}}  \text{ on }\widetilde{\omega}\setminus \omega &\textit{ (PFC correction)}
   \end{array}\right. \\
   \\
    \circled{5}   \text{ get from }  \prescript{s+1}{}{\psi_\text{PFC}} \text{ the corresponding amplitudes on } \omega \text{ and interpolate them } \quad\textit{ (local PFC interpolation)}\\
    \\
        \circled{6}    \text{ correct }\left\{ \prescript{s+1}{}{\eta_m}\right\}_{m=1}^3, \prescript{s+1}{}{\psi_0} \text{ by the corresponding amplitudes from the PFC on } \omega \textit{ (APFC correction)}
   \end{array} \right.\\
  \label{eq:hybrid}
\end{equation}

Step $\circled{1}$ and $\circled{2}$ correspond to the classical updating schemes,  Eqs.~\eqref{eq:apfc_fft} and \eqref{eq:conv}. By an appropriate definition of $\omega$ and $\widetilde{\omega}$, step $\circled{4}$ and $\circled{6}$ realize the coupling between the PFC and APFC within the hybrid-PFC model.
Since these are solved on different numerical grids with different grid resolutions, an interpolation on a common grid is needed. The interpolation procedure will negatively affect the computational costs, which is why it needs to be done locally on $\widetilde{\omega}$ (step $\circled{3}$ and $\circled{5}$). 
We point out that the local PFC updates have to be done with the \convalg~\eqref{eq:conv} with linear convolutions. This algorithm is indeed crucial in our hybrid setting because the \fftalg~\eqref{eq:fft} would require a global (and therefore costly) update of $\prescript{s+1}{}{\psi_\text{PFC}}$, preventing a faster computation than the original PFC model. 
By omitting steps $\circled{5}$ and $\circled{6}$ in Eq.~\eqref{eq:hybrid}, one obtains a simplified approach that passes information from the APFC to the PFC but not vice versa. In the following, we further illustrate the approach in practical settings and showcase the capabilities of the full and simplified hybrid-PFC models through numerical benchmark simulations, for which we provide run-time and convergence studies.

\subsection{Benchmarks of the hybrid-PFC model against PFC and APFC models}

\begin{figure}[h!]
\centering
\includegraphics[width=\textwidth]{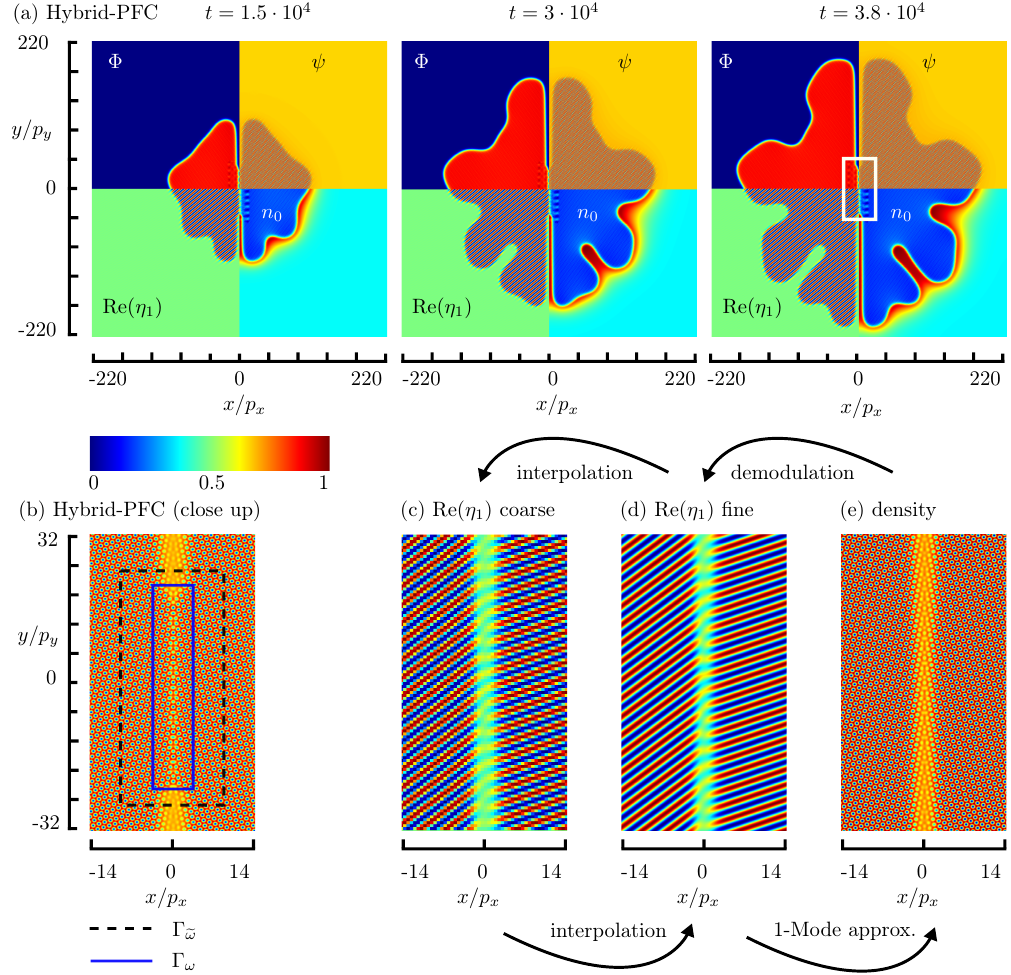}
 \caption{Formation of a GB between two grains of different orientations during solidification as an example for the hybrid-PFC model. (a) Snapshots of the density $\psi$ (top right), order parameter $\Phi$ (top left), the real part of amplitude $\eta_1$ (bottom left), and the local average density $\psi_0$ (bottom right) at different time steps. For illustration purposes, we interpolated  $\Phi$, $\mathrm{Re}(\eta_1)$  and $\psi_0$ on a fine grid and normalized all plotted quantities from 0 to 1. Values of $\psi$, $\Phi$, $\psi_0$ and $\mathrm{Re}(\eta_1)$ vary within the following ranges: $-0.08\leq \psi\leq 1.32$, $0\leq \Phi\leq 0.12$, $0.82 \leq \psi_0 \leq 0.89$ and $-0.14\leq \mathrm{Re}(\eta_1) \leq 0.14$. (b) Magnification of the region denoted with the white rectangle in panel (a) at $t=3.8\cdot 10^4$. It shows explicitly the domains $\omega$ and $\widetilde{\omega}$. (c) $\mathrm{Re}(\eta_1)$ on the coarse APFC grid. (d) $\mathrm{Re}(\eta_1)$ on a fine locally interpolated grid. (e) Reconstructed density from amplitudes in panel (d) via Eq.~\eqref{eq:1mode} on a fine grid. Subplots (c)-(e) illustrate the interpolation between the coarse APFC and the fine PFC grid. Further explanations are given in the main text. Lengths are scaled with the atomic spacings along the $x$- and $y$-axis, $p_x$ and $p_y$.}
\label{fig:hybrid_scetch}
\end{figure}

We illustrate the proposed hybrid-PFC approach by first considering a setting similar to Fig.~\ref{fig:pfc_apfc} with parameters reported in Table~\ref{tab:numerics}. We prepare two solid seeds with a rotation angle of $\pm 19^\circ$ in an undercooled melt, see Fig.~\ref{fig:hybrid_scetch}(a) where we show the resulting density field $\psi$ (obtained by setting $\left. \psi \right|_{\Omega\setminus \omega} = \psi_\text{APFC}$ and $\left. \psi \right|_{\omega} = \psi_\text{PFC}$), phase-field $\Phi$, the real part of $\eta_1$ and local average density $\psi_0$ at different time steps during growth. The imposed rotation leads to complex amplitudes $ \left\{ \prescript{s+1}{}{\eta_m}\right\}_{m=1}^3$ whose phase encodes the lattice deformation \cite{salvalaglio2022coarse}, see Fig.~\ref{fig:hybrid_scetch}(a). Besides the growth of two dendrite-like structures, a GB forms between the two grains in the central region; see also a magnification in Fig.~\ref{fig:hybrid_scetch}(b). The boundaries of the two domains $\omega=[-3.4,3.4]p_x\times [-23.9,20.7]p_y$ and $\widetilde{\omega}=[-8.3,8.3]p_x\times [-28.6,25.5]p_y$ as defined within the scheme~\eqref{eq:hybrid} are shown in Fig.~\ref{fig:hybrid_scetch}(b). The transition from Fig.~\ref{fig:hybrid_scetch}(c) to Fig.~\ref{fig:hybrid_scetch}(d) shows the local interpolation of $\prescript{s+1}{}{\eta_1}$ from the coarse APFC grid to the fine PFC grid. In \ref{sec:interpol}, further details about this numerical interpolation are reported. Note that $\prescript{s+1}{}{\eta_2}$, $\prescript{s+1}{}{\eta_3}$ and $\prescript{s+1}{}{\psi_0}$ are interpolated as well to compute the corresponding density $\prescript{s+1}{}{\psi_\text{APFC}}$ via the 1-Mode approximation~\eqref{eq:1mode}, see Fig.~\ref{fig:hybrid_scetch}(e). The transition \ref{fig:hybrid_scetch}(c)$\rightarrow$\ref{fig:hybrid_scetch}(d)$\rightarrow$\ref{fig:hybrid_scetch}(e) corresponds to step $\circled{3}$ in~\eqref{eq:hybrid}. In the region $\widetilde{\omega}\setminus\omega$ the PFC density $\prescript{s+1}{}{\psi_\text{PFC}}$ is directly corrected by the resulting density $\prescript{s+1}{}{\psi_\text{APFC}}$ from \ref{fig:hybrid_scetch}(e), corresponding to step $\circled{4}$ in Eq.~\eqref{eq:hybrid} ($\left. \prescript{s+1}{}{\psi_\text{PFC}}\right|_{\widetilde{\omega}\setminus \omega} \coloneqq  \left. \prescript{s+1}{}{\psi_\text{APFC}}\right|_{\widetilde{\omega}\setminus \omega}$). Inverting the steps above corresponds to $\circled{5}$ in~\eqref{eq:hybrid}. In particular, by coarse-graining the PFC density $\prescript{s+1}{}{\psi_\text{PFC}}$ via demodulation (further details in \ref{sec:interpol}), amplitudes are computed on the fine PFC grid, see transition \ref{fig:hybrid_scetch}(e)$\rightarrow$\ref{fig:hybrid_scetch}(d) and then extracted on the coarse APFC grid; see \ref{fig:hybrid_scetch}(d)$\rightarrow$\ref{fig:hybrid_scetch}(c). Finally, the APFC amplitudes $\left\{ \prescript{s+1}{}{\eta_m}\right\}_{m=1}^3$ and $\psi_0$ are directly corrected by those extracted from the PFC density on $\omega$, i.e. step $\circled{6}$ in~\eqref{eq:hybrid}.
 We point out that the hybrid-PFC model allows for a consistent coupling between the APFC and the PFC formulations. Indeed, both the density field for a local PFC update $\prescript{s+1}{}{\psi_\text{PFC}}$ as well as the local average density and amplitudes for a global APFC update $\psi_0, \left\{ \prescript{s+1}{}{\eta_m}\right\}_{m=1}^3$  show smooth transitions at the boundaries $\Gamma_\omega$ and $\Gamma_{\widetilde{\omega}}$ during simulation, see Fig.~\ref{fig:hybrid_scetch}(b),(c). Further evidence of the proper approximation of the PFC model by the hybrid-PFC model is reported below.
 
\begin{figure}[h!]
\centering
\includegraphics[width=\textwidth]{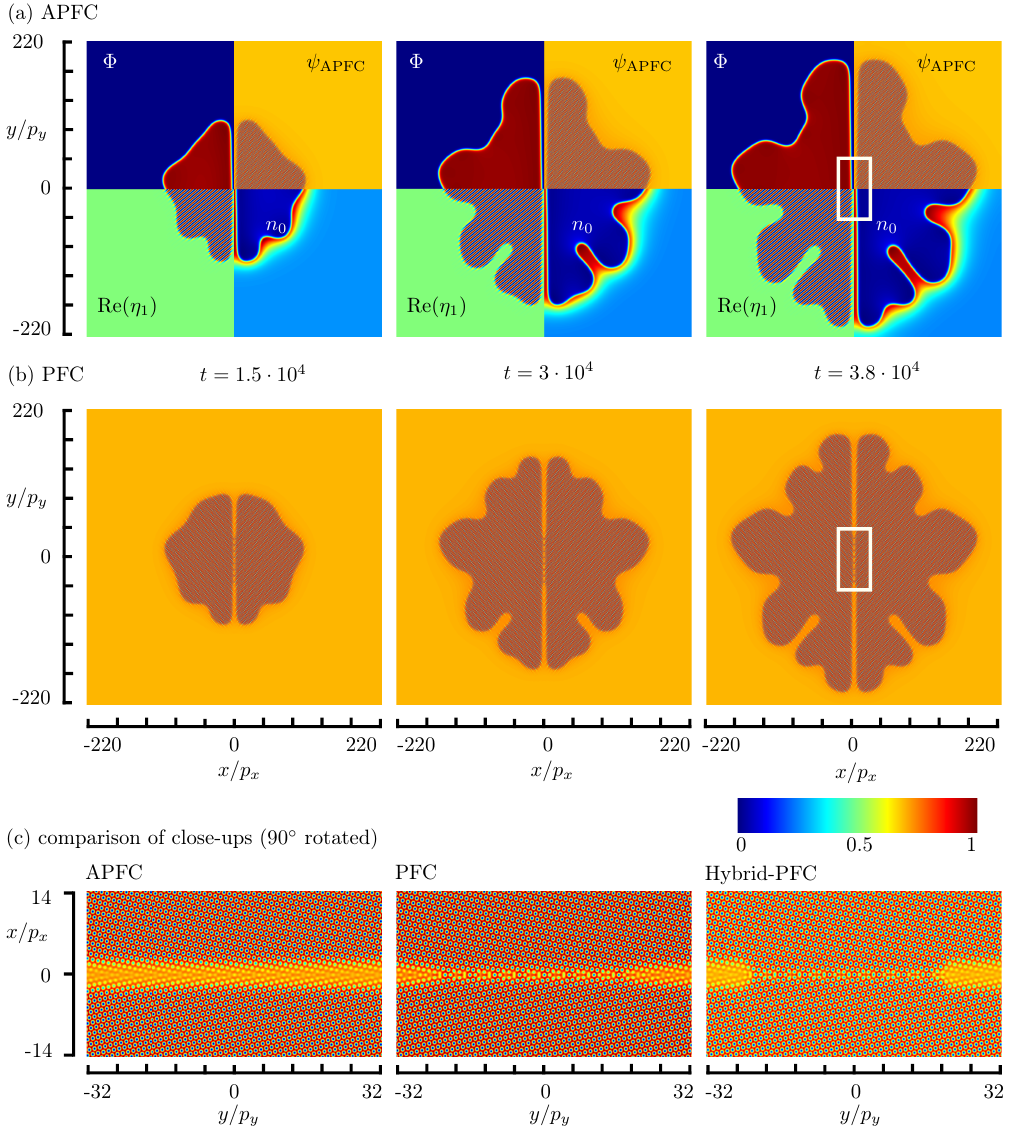}
\caption{Comparison of PFC, APFC, and hybrid-PFC model for the solidification of two solid seeds with large relative rotation angle as Fig.~\ref{fig:hybrid_scetch}. (a) Reconstructed density, $\Phi$, the real part of the amplitude $\eta_1$, and the local average density $\psi_0$ at different, representative times for the APFC model. (b) Density obtained by the PFC model at different, representative times. For illustration purposes, we snormalized all plotted quantities from 0 to 1. Values of $\psi_\text{APFC}$, $\Phi$, $\mathrm{Re}(\eta_1)$, $\psi_0$  and $\psi_\text{PFC}$ vary within the following ranges: $0.05\leq \psi_\text{APFC}\leq 1.23$, $0\leq \Phi\leq 0.10$, $-0.12\leq \mathrm{Re}(\eta_1)\leq 0.12$, $0.83 \leq \psi_0 \leq 0.89$ and $0.03\leq \psi_\text{PFC}\leq 1.21$ (c) Magnification the region denoted with the white rectangles in (a), (b), and Fig.~\ref{fig:hybrid_scetch}(b). Lengths are scaled with the atomic spacings along the $x$- and $y$-axis, $p_x$ and $p_y$.}
\label{fig:beats}
\end{figure}

 In Fig.~\ref{fig:beats}, the hybrid-PFC model is compared with the simulations of the APFC and the PFC model for the settings illustrated in Fig.~\ref{fig:hybrid_scetch}. In particular, Fig.~\ref{fig:beats}(a) shows snapshots of the reconstructed density, $\Phi$, $\mathrm{Re}(\eta_1)$ and $\psi_0$ resulting from an APFC simulation. Fig.~\ref{fig:beats}(b) shows the density obtained by PFC simulations. In Fig.~\ref{fig:beats}(c), we show an enlarged view of the region where the two growing seeds merge at $t=3.5\cdot10^4$ (marked by the white rectangle in panels (a) and (b)). With the APFC model, this region features a liquid layer, and the two growing crystals remain separated (see Fig.~\ref{fig:beats}(a,c)). This is a spurious result due to the poor description of largely rotated crystals by the APFC model \cite{salvalaglio2022coarse,huter2017modelling,spatschek2010amplitude}. Indeed, in its original formulation, the PFC model predicts the formation of a GB (see also Fig.~\ref{fig:GB}). The hybrid-PFC formulation can model the same GB structure as in the PFC model, see Fig.~\ref{fig:beats}(c), although a coarse APFC resolution is used far away from the GB. Importantly, while the PFC model requires a run-time of $\approx 6.3\cdot 10^4$ seconds, the Hybrid model reduces the run-time by approximately half an order of magnitude when considering the simplified formulation and the SVD as illustrated in the Appendix~\ref{fig:svd}(b) and~\ref{fig:omega}.
 
\begin{figure}[h!]
\centering
\includegraphics[width=\textwidth]{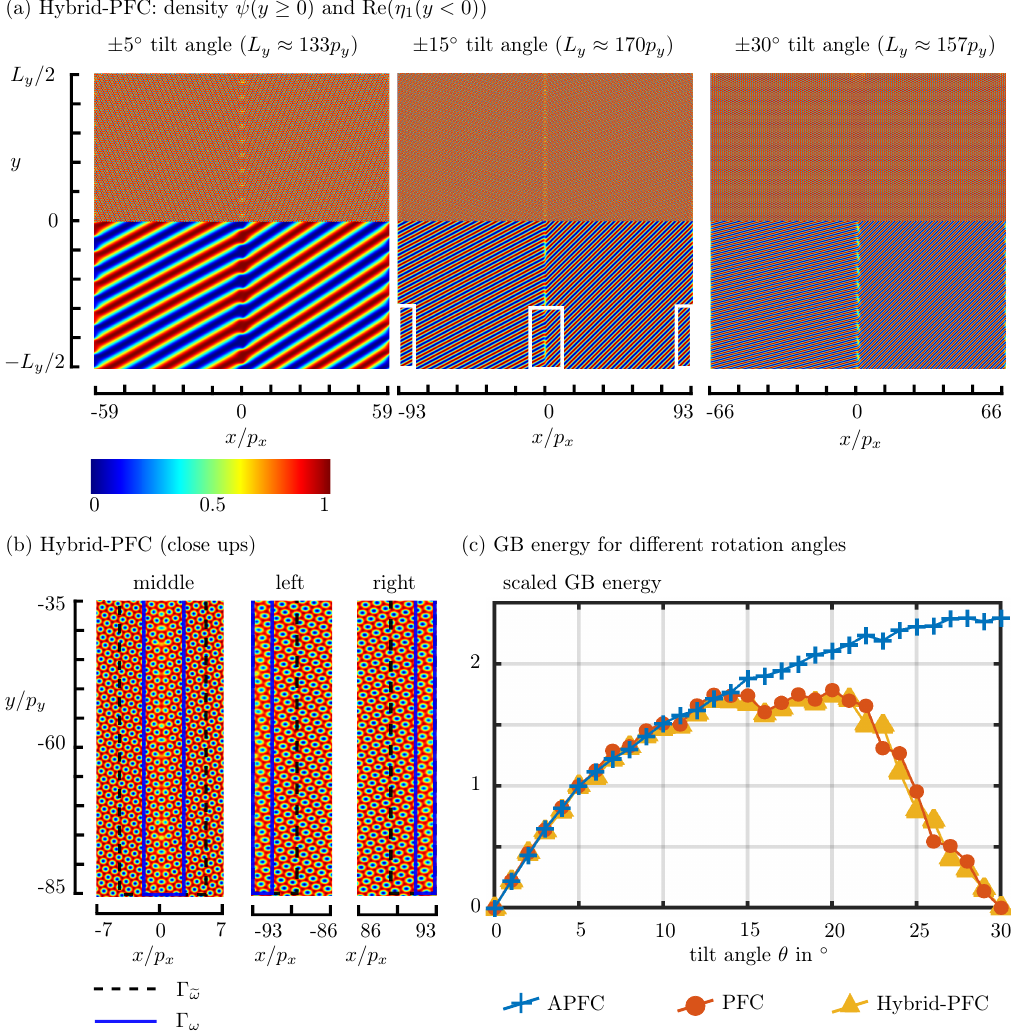}
\caption{GB energy for different rotation angles. (a) $\psi$ and real part of amplitudes $\eta_1$ for tilt angles $\pm 5^\circ$, $\pm 15^\circ$ and $\pm 30^\circ$ obtained by the hybrid-PFC model. In every subplot the upper half shows the density field  $\psi(y\geq 0)$, while the lower half shows $\mathrm{Re}(\eta_1(y<0))$. (b) Magnification of the white-colored domains from (a), including parts of the domains $\omega$ and $\widetilde{\omega}$ are shown. (c) GB energy for different tilt angles obtained by the APFC, PFC, and hybrid-PFC models. For illustration purposes, we scale all the plots to have the same energy for $\theta=5^{\circ}$ \cite{hirvonen2016multiscale}. Values of $\psi$ and $\mathrm{Re}(\eta_1)$ vary in  the following intervals: $-0.06\leq \psi \leq 1.27$ and $-0.15\leq \mathrm{Re}(\eta_1)\leq 0.15$.  Lengths are scaled with the atomic spacings along the $x$- and $y$-axis, $p_x$ and $p_y$.}
\label{fig:energy}
\end{figure}

The second benchmark, further addressing comparisons of models and quantitative aspects, focuses on GB structures and energies. Symmetric GBs forming between crystals with underlying triangular crystal symmetries are considered. The GB energy as a function of the tilt angle $2\theta$ is computed, similar to settings explored, e.g., in Refs.~\cite{hirvonen2016multiscale,salvalaglio2017controlling,mellenthin2008phase}. In particular, we set a rectangular domain $\Omega=[-L_x/2,L_x/2]\times [-L_y/2,L_y/2]$ and impose a rotation of $\pm \theta$ on the left and right crystal. As a result, vertical GBs are formed in the middle and at the boundary of the domain $\Omega$, see Fig.~\ref{fig:energy}(a). We vary the tilt angle $\theta \in [0^\circ, \,30^\circ]$, let the initial condition relax ($t=5\cdot 10^3$), and measure the GB energy per unit length. Model parameters are reported in Table~\ref{tab:numerics}.
In Fig.~\ref{fig:energy}(a), we showcase  the density fields $\psi$ for the hybrid-PFC model (obtained by setting $\left. \psi \right|_{\Omega\setminus \omega} = \psi_\text{APFC}$ and $\left. \psi \right|_{\omega} = \psi_\text{PFC}$) as well as the real part of corresponding amplitudes $\eta_1$ for different rotation angles ($\pm 5^\circ,\,\pm 15^\circ, \pm 30^\circ$ ). Similar to Fig.~\ref{fig:GB}, $\Omega$ is set to a multiple of the period of $\psi$ along $x$- and $y$-directions as imposed by the rotated lattice vectors which depend on $\theta$, to impose a strain-free system~\cite{mellenthin2008phase,blixt2021evaluation}. 
For our hybrid-PFC model, we choose the domain $\omega=\bigcup_{i=1}^3 \omega_i$ and $\widetilde{\omega}= \bigcup_{i=1}^3 \widetilde{\omega}_i,$ for the local PFC update along the GBs with
\begin{equation}
    \begin{split}
        & \begin{array}{ll} \left. \begin{array}{ll}
        \omega_1=[-11.7p_x,11.7p_x]\times [-L_y/2,L_y/2],\\
        \widetilde{\omega}_1=[-13.8p_x,13.8p_x]\times [-L_y/2,L_y/2]
        \end{array} \right\}& \textit{middle GB,} \end{array}\\
         & \begin{array}{ll} \left. \begin{array}{ll}
       \omega_2= [-L_x/2,-L_x/2+11.7p_x]\times [-L_y/2,L_y/2],\\
       \widetilde{\omega}_2= [-L_x/2,-L_x/2+13.8p_x]\times [-L_y/2,L_y/2]
        \end{array} \right\}& \textit{left domain boundary,} \end{array}\\
          & \begin{array}{ll} \left. \begin{array}{ll}
       \omega_3 = [L_x/2-11.7p_x,L_x/2]\times [-L_y/2,L_y/2],\\
        \widetilde{\omega}_3 = [L_x/2-13.8 p_x,L_x/2]\times [-L_y/2,L_y/2]
        \end{array} \right\}& \textit{right domain boundary.} \end{array}\\
    \end{split}
\end{equation}
In Fig.~\ref{fig:energy}(b), a magnification of the area within the white rectangle in  Fig.~\ref{fig:energy}(a) is shown, together with representative parts of the domains $\omega$ and $\widetilde{\omega}$ for $\theta = 15^\circ$. In Fig.~\ref{fig:energy}(c), we compare the GB energy in a region along the central GB computed from the solution of the PFC and APFC models via Eq.~\eqref{eq:F1} and~\eqref{eq:apfc_energy}, respectively. We also show the GB energy computed for the hybrid PFC, evaluated as for the PFC model via Eq.~\eqref{eq:F1}. To compare these quantities, we normalize the curves for $\theta = 5^\circ$, where both PFC and APFC are expected to perform well. Note that the normalization allows for comparisons without reparametrization of the free energies as commonly adopted when comparing these models \cite{hirvonen2016multiscale}. For $\theta\leq 10^\circ$, we obtain that the description of all the considered approaches is de facto equivalent, evidence motivating the usage of the APFC model for small tilts (i.e. when isolated dislocations only form. For relatively large tilt angles, $\theta \geq 15^\circ$, the APFC deviates from the results of the PFC model until predicting an unphysical boundary for large tilts where the effective misorientation actually decreases (and so does the energy predicted by the more accurate PFC model) owing to the symmetry of the considered triangular lattice. However, the hybrid-PFC model matches the result of the PFC model almost perfectly. We note that modifications of the APFC model or settings for the APFC simulations may lead to a decrease of the GB energy of high angle GBs~\cite{salvalaglio2017controlling,hirvonen2016multiscale,mellenthin2008phase,bervcivc2020enabling}. They however require to either modify the models/equations or perform simulations with different settings (e.g. different definition of the initial conditions) to obtain one GB energy curve. Here we considering a consistent coupling of standard methods, which is then amenable to include all recent and future developments of PFC and APFC models straightforwardly. Besides, APFC model only cannot describe fine details as small local minima (e.g., at $\theta \sim 16^\circ$).
 
Overall, it is thus shown that the hybrid-PFC formulation allows for computing GB energy in good quantitative agreement with the PFC model. Even though the domains with local PFC accuracy within the hybrid-PFC framework are here chosen relatively large compared to the whole domain, owing the specific geometry and to highlight the quantitative agreement with the PFC model that can be achieved, if needed, 
approximately half of the computational costs can be saved when considering the simplified hybrid-PFC model with an SVD as described in the Appendix~\ref{fig:svd},~\ref{fig:refine} (for $\theta=15^\circ$, the PFC model requires a run-time of $\approx 790$s, while the hybrid-PFC requires a run-time of $\approx 420$s). Moreover, no optimization of the domains is considered, which can be, however, envisaged via the composition of domains with simple rectangular shapes leveraging the same algorithms presented here. This will pave the way for further significant improvement in performance and further applications.

\section{Conclusion}
\label{sec:conclusions}
We introduced a hybrid-PFC model, allowing for multi-scale simulations with local PFC accuracy in regions where microscopic resolution is needed while exploiting the coarse-grained description conveyed by the APFC model elsewhere. 

An alternative real-space implementation of the PFC equation leveraging concepts of the Fourier pseudo-spectral method but exploiting local (real-space) convolutions has been introduced. This method delivers results equivalent to the state-of-the-art scheme based on the fast Fourier transform algorithm but allows for local updates needed for the proposed hybrid-PFC method. Furthermore, the numerical method based on real-space convolutions allows for considering boundary conditions other than simply periodic, beyond the capabilities of most Fourier spectral methods.

The proposed hybrid-PFC model solves the APFC model globally on a coarse grid, while the PFC provides local corrections via the convolution algorithm mentioned above. Through numerical simulations, we showcase the model's capabilities and benchmark it against classical PFC and APFC models. It is shown that the hybrid-PFC framework may lead to high accuracy in approaching the results of PFC simulations while allowing for fast run times, two aspects that are not accessible simultaneously when considering the classical, uncoupled models. This of course holds true when high accuracy is needed in some limited regions in the domain although the PFC model should be considered if these regions extend over most of the domain.

This work sets the ground for efficient multi-scale simulations of crystalline materials, retaining microscopic details locally and meso- to macroscopic length scales. Perspective extensions include implementing algorithmically efficient determination of the region where to solve PFC equations within the hybrid-PFC model based on the quantity of interest in an adaptive fashion. This may include, for instance, criteria based on variation of $\Phi$ to detect the nucleation of defects or the formation and evolution of interfaces.
Also, three-dimensional settings can be built on the same concepts proposed here while being amenable to PFC and APFC simulations. Real-space methods that proved powerful to handle adaptivity for APFC simulations, e.g., in Refs.~\cite{bervcivc2018adaptive,Praetorius2019}, can also be combined with the concepts presented here. They would relax the need for the convolution algorithm but would still exploit the other proposed steps to couple PFC and APFC within the hybrid-PFC model. However, we remark that the hybrid-PFC model exploits uniform -but different- spatial discretizations, solving the APFC and PFC models separately.

We finally note that since the APFC model is based on free energy derived from the PFC model, extensions of the minimal models considered can be readily devised by considering the corresponding formulations and proceed via the exchange of information via the interpolations of the density field(s) and complex amplitudes. Accordingly, the dynamic influenced by other physical contributions can be inspected, while interpolations between quantities and the numerical algorithm, as seen in this work, can be applied straightforwardly.


\addcontentsline{toc}{section}{CRediT authorship contribution statement}
\section*{CRediT authorship contribution statement}
\textbf{Maik Punke:} Conceptualization, Data curation, Formal analysis, Investigation, Methodology, Software, Validation, Visualization, Writing – original draft, Writing – review and editing. \textbf{Marco Salvalaglio:} Conceptualization, 
Formal analysis, Funding acquisition, Methodology, Project administration, Resources, 
Supervision, Validation, 
Writing – review and editing.
\addcontentsline{toc}{section}{Declaration of competing interest}
\section*{Declaration of competing interest}

The authors declare that they have no known competing financial interests or personal relationships that could have appeared to influence the work reported in this paper.

\addcontentsline{toc}{section}{Data availability}
\section*{Data availability}

The data that support the findings of this study will be made openly available on suitable repositories in the final version.

\addcontentsline{toc}{section}{Acknowledgments}
\section*{Acknowledgments}
The authors gratefully acknowledge support from the German Research Foundation (DFG) under Grant SA4032/2-1 (Project No. 447241406) and the computing time made available to them on the high-performance computer at the NHR Center of TU Dresden. This center is jointly supported by the Federal Ministry of Education and Research and the state governments participating in the NHR (\href{www.nhr-verein.de/unsere-partner}{www.nhr-verein.de/unsere-partner}).

\vspace{\baselineskip}

\addcontentsline{toc}{section}{References}
\bibliographystyle{elsarticle-harv} 
\bibliography{references}

\appendix

\section{Simulation Parameters}
\label{sec:appendixTable}

\begin{table}[h!]
    \centering
    \resizebox{\textwidth}{!}{
    \begin{tabular}{llllllllllllllll}
    \hline
         Figure&&Model&& \multicolumn{7}{l}{Model Parameters}&&\multicolumn{3}{l}{Numerical Parameters}\\ \cline{5-11}  \cline{13-15} 
     & &&& M&$\lambda$ &$\kappa$  &$\Psi _0$& $\delta$ & $\Omega$ in UC &t  &&$\Delta x$&$\Delta y$&$\Delta t$ \\
        \hline 
        \ref{fig:GB}, \ref{fig:svd}(a)&&PFC&&1 &0.23  & 1/3 &-0.2&0 &$\approx 1342 \times 16$ &$10^3$&&0.63&0.63&0.1\\
         \hline
          \ref{fig:random}& &&& &0.6  & 0.46 & 0.849&1&$10\times 10$&$2\cdot 10^4$ &&0.73&0.63&\\
                   \hline
         \ref{fig:pfc_apfc} &&&&0.66 &&&&&$\approx220\times 255$&$5.3\cdot 10^4$&&1.33 &1.20 &\\
       &&APFC&  &-&&&&&&&&8.90 &7.70&\\
       \hline
         \ref{fig:refine} &&PFC, APFC&& 0.66&&&&&&&&0.50\dots 8.90 &0.43 \dots 7.70 &\\
      \hline
      \ref{fig:hybrid_scetch}, \ref{fig:svd}(b), \ref{fig:omega}(a) && hybrid-PFC&& &&&&&&$3.8\cdot 10^4$&&4.00&3.50&\\
      & & &&&&&&&&&&0.50 (on $\widetilde{\omega}$)&0.43 (on $\widetilde{\omega}$ )&\\
       \hline
     \ref{fig:beats} &&PFC&& &&&&&&&&0.50&0.43&\\
      && APFC&&- &&&&&&&&&\\
      \hline
    \ref{fig:energy}, \ref{fig:svd}(c), \ref{fig:omega}(b) &&PFC&&0.66&&&0.82&&$\approx$ 186 $\times$170&$5\cdot 10^3$&&0.50&0.43&1\\
      & &APFC&&- &&&&&&&&&&\\
       &&hybrid-PFC&&&&&&&&&&3.70&3.00&\\
       &&&&&&&&&&&&0.50 (on $\widetilde{\omega}$)&0.43 (on $\widetilde{\omega}$)&\\
                 \hline  \end{tabular}}
    \caption{Model and numerical parameters 
    for all the simulations reported in this paper. For the setup of Fig.~\ref{fig:GB} and \ref{fig:svd}(a) we model underlying square crystal symmetry with $Q=2$, $q_1=1$ and $q_2=\sqrt{2}$. The other setups consider an underlying triangular crystal symmetry with  $Q=1$ and $q_1=1$. These quantities enter the $\mathcal{L}$ operator in Eq.~\eqref{eq:F1}. Empty table entries read as the row above.}
    \label{tab:numerics}
\end{table}

\section{Implementation details and auxiliary numerical studies}
\label{sec:appendix}
In this appendix, further details about an efficient implementation and numerical approximations of the \convalg~\eqref{eq:conv} explained in section~\ref{sec:conv} are given. For simplicity, we start with a 1D setup and showcase the extensions to higher dimensions. 
Furthermore, we present refinement and domain studies for selected simulation results presented in the main text, see~\ref{sec:refinement_domain}. Finally, we summarize the Fourier zero-padding method as an interpolation technique used for illustrating purposes as well as for our hybrid-PFC model~\ref{sec:interpol}.

\subsection{Efficient Implementation of the convolution algorithm}
\label{sec:conv}
\subsubsection{Linear convolutions with local updates}
\label{sec:linconv}
The central part of the linear (or direct) convolution of two given vectors $A\in \mathbb{R}^m$, $B\in \mathbb{R}^n$  can be defined as
\begin{equation}
    \begin{split}
        (A*B)_k=\sum_{i=\max(1,k+1-\lfloor n/2\rfloor )}^{\min(k+\lfloor n/2 \rfloor ,m)}A_iB_{\lfloor n/2 \rfloor-k+i}, \quad \forall k=1,\dots, m
    \end{split}
    \label{eq:convAlg}
\end{equation}
with  $A_i$ being the $i^\text{th}$  entry of $A$ and $\lfloor \cdot \rfloor$ denoting the floor function. The convolution~\eqref{eq:convAlg} can be seen as the replacement of $A$ by its weighted mean with $B$ in a surrounding of length $m$. In~\eqref{eq:convAlg}, $B$ is partially weighted with zero for $k< \lfloor n/2\rfloor$ (the entries $B_{\lfloor n/2 \rfloor+k} \dots B_{n}$ are multiplied with zero) and $k> m-\lfloor n/2\rfloor$ (the entries $ B_1 \dots B_{k-m+1+\lfloor n/2 \rfloor}$ are multiplied with zero). Within the \convalgnospace~\eqref{eq:conv}, this is used to implement DBC leveraging the possibility to update only some of the terms in the sum in Eq.~\eqref{eq:convAlg} and fix others to some values.

In brief, Eq.~\eqref{eq:convAlg} can be modified to compute a local convolution of $A$ and $B$ considering a splitting
\begin{equation}
    (A*B)_k=(\widetilde{A}*B)_{k-q+1}+(Z*B)_k,\quad \forall k\in K,
    \label{eq:split}
\end{equation}
for $1\leq q\leq r\leq m$, $\widetilde{A}\in \mathbb{R}^{r-q+1}$ with $\widetilde{A}_k=A_{k+q-1}$, $Z\in \mathbb{R}^m$ with $Z_k=0\,  \forall\, k\in K$ and $Z_k=A_k$ else,
and $K=\lbrace q,\dots ,r \rbrace$ the indices of interest over which the evaluation of the convolution is restricted.
\eqref{eq:split} is convenient w.r.t \eqref{eq:convAlg} if $(A*B)_k$ needs to be computed multiple times for changing values $A_k$ but fixed indices of interest $k\in K$, e.g. within an iterative scheme. In the \convalgnospace~\eqref{eq:conv}, involving two convolutions, one can set first $A=\prescript{s}{}{A}=-\delta/2\prescript{s}{}{\psi^2} +\prescript{s}{} \psi^3/3\in \mathbb{R}^m$, $B=Q\in \mathbb{R}^n$ and $n=m$ with $s$ labeling the solution at time $t=s\Delta t$ as in the main text. Then the same concept can be applied to the second convolution now reading $\left[\prescript{s}{}{\psi}+\prescript{s}{}{(A*B)}\right]* R$. As a result, at every time step $s$, a local density update for the indices of interest $k\in K$ requires the computation of two convolutions at $r-q+1$ points instead of two convolutions at $n$ points. The weight of $B$ and $R$ by zero for $k< \lfloor n/2\rfloor$ and $k> m-\lfloor n/2\rfloor$ holding true for every time step $s$ directly enforce a DBC. We recall that it is possible to model the DBC with modifications of the \fftalgnospace~\eqref{eq:fft} too. A possibility is to pad $\prescript{s}{}{\psi}\in \mathbb{R}^m$ with $\lfloor m/2\rfloor$ zeros along each boundary at every time step: $[0,\dots,0,\prescript{s}{}{\psi_1},\dots,\prescript{s}{}{\psi_m},0,\dots,0]\in \mathbb{R}^{m+2\lfloor m/2\rfloor}$
Another possibility is to use discrete sine-transforms~\cite{schumann1988fast}.

The convolution defined in Eq.~\eqref{eq:convAlg} can be extended to higher dimensions, e.g. in a 2D setting with $A\in \mathbb{R}^{n_1\times n_2}$ and $B\in \mathbb{R}^{m_1\times m_2}$
\begin{equation}
    \begin{split}
        (A*B)_{k,l}=\sum_{i=\max(1, k+1-\lfloor n_1/2 \rfloor )}^{\min(k+\lfloor n_1/2 \rfloor, m_1)}\sum_{j=\max(1, l+1-\lfloor n_2/2 \rfloor )}^{\min(l+ \lfloor n_2/2 \rfloor, m_2)}A_{i,j} B_{\lfloor n_1/2 \rfloor -k+i, \lfloor n_2/2 \rfloor -l+j}, \quad \forall k=1,\dots, n_1 \;\text{and }  l=1,\dots, n_2
    \end{split}
    \label{eq:convAlg2}
\end{equation}
with $A_{k,l}$ referring to the $k^\mathrm{th}$ row and $l^\mathrm{th}$ column of the matrix $A$. The splitting~\eqref{eq:split} can be adapted straightforwardly to a 2D setting, e.g. for the setup in Fig.~\ref{fig:GB} the discretization of $[-\Omega_x/2,\Omega_x/2]\times [-L_y/2,L_y/2]$    gives the indices of interest $K$. For the hybrid-PFC setups illustrated in Fig.~\ref{fig:hybrid_scetch} and Fig.~\ref{fig:energy}, the discretization of $\omega$ gives the indices of interest $K$ over which convolutions are computed and the solution is thus updated using the chosen integration scheme. 

Most numerical libraries include 1D~\eqref{eq:convAlg}, 2D~\eqref{eq:convAlg2} and higher dimensional linear convolutions due to their wide range of applications, e.g. image and signal processing~\cite{keys1981cubic,damelin2012mathematics}, convolutional neural networks~\cite{yang2017application,albawi2017understanding} as well as numerical discretizations within the finite-difference method (FDM) framework~\cite{fornberg1975fourier,fornberg1987pseudospectral}. 


\subsubsection{Cyclic convolutions}
\label{sec:cycconv}
The implementation of a cyclic convolution of two vectors $A\in \mathbb{R}^m$ and $B\in\mathbb{R}^n$ which is not based on an fft algorithm can be obtained by linear convolutions \eqref{eq:convAlg}. A simple way is to mirror the vector $A$ via $[A_2,\dots,A_m,A_1,\dots,A_m]$  and use a linear convolution $C\coloneq [A_2,\dots,A_m,A_1,\dots,A_m] *B\in \mathbb{R}^{2m-1}$. The cyclic convolution can now be written as $A\raisebox{.5pt}{\textcircled{\raisebox{-2.0pt} {*}}}B= [C_m,\dots, C_{\left \lfloor{3/2m}\right \rfloor}, C_{\left \lfloor{m/2}\right \rfloor +1},\dots, C_{m-1}]$. Similar to~\eqref{eq:split} and~\eqref{eq:convAlg2}, cyclic convolutions can be computed locally and extended to higher dimensions. Contrary to~\eqref{eq:convAlg}, $B$ is no longer partially weighted with zero but instead with the periodic image of $A$ for $k< \lfloor n/2\rfloor$ and $k> m-\lfloor n/2\rfloor$, enforcing PBC. For the \fftalgnospace~\eqref{eq:fft} no further modifications are necessary to impose PBC.

 \subsubsection{Sparse convolutions}
Convolutions of sparse  arrays are in general computationally cheaper than convolutions of  non-sparse arrays. The two convolutions needed for the updating scheme \convalgnospace~\eqref{eq:conv}  can be decomposed into  
\begin{equation*}
\begin{split}
 \left[\prescript{s}{}{\psi} +  \left(-\dfrac{\delta}{2} \prescript{s}{}{\psi^2} + \dfrac{ \prescript{s}{}{\psi^3}}{3}\right)\raisebox{.5pt}{\textcircled{\raisebox{-2.5pt} {*}}}Q\right]\raisebox{.5pt}{\textcircled{\raisebox{-2.5pt} {*}}}R &=  \left[\prescript{s}{}{\psi}-\Psi_0 + \left(-\dfrac{\delta}{2} (\prescript{s}{}{\psi -\Psi_0)}^2 + \dfrac{(\prescript{s}{}{\psi-\Psi_0)}^3}{3}\right)\raisebox{.5pt}{\textcircled{\raisebox{-2.5pt} {*}}}Q\right]\raisebox{.5pt}{\textcircled{\raisebox{-2.5pt} {*}}}R +\Psi_0
       \end{split}
        \end{equation*}
by using the linearity of the convolution. Especially if a locally constant density with $\Psi_0\neq0$ is modeled (liquid crystal phase), an implementation of the right-hand side of the equation above shows reduced computational costs compared to an implementation of the left-hand side. Local updates~\eqref{eq:split}, cyclic convolutions~\eqref{sec:cycconv} as well as the extension to higher dimensions~\eqref{eq:convAlg2} can be applied straightforwardly.
\subsubsection{Lower dimensional decompositions}
\label{sec:svd}

For 2D or higher dimensional convolutions, the computational costs can be further reduced by decomposing them into 1D convolutions. For instance, the 2D field $Q=\Delta t M\widehat{k^2}$ defined for the \convalgnospace~\eqref{eq:conv} can be written as the sum  $Q=\Delta t M(\widehat{k_1^2}+\widehat{k_2^2})$ with vectors $k_1$ and $k_2$. Any convolution $(\cdot)*Q$ can therefore be computed by convolving each column of $(\cdot)$ with $\Delta t M \widehat{k_2^2}$ and then convolving each row of the resulting two-dimensional matrix with $\Delta t M\widehat{k_1^2}$. 

Not every 2D field can be simply decomposed as $Q$. For instance, this is the case for $R$ entering the \convalgnospace~\eqref{eq:conv}. However, one can extend the concept mentioned above by the so-called singular value decomposition (SVD). 
The SVD can be defined as follows: let $R\in \mathbb{R}^{n_1\times n_2}$, then there are orthogonal matrices $U\in O(n_1), V\in O(n_2)$ and an $n_1\times n_2$ diagonal matrix $S =\mathrm{diag(\sigma_1,\dots \sigma_{\min (n_1,n_2)}})$ with singular values $\sigma_1 \geq\sigma_2 \geq\dots \geq \sigma_{\min (n_1,n_2)}\geq 0$ such that we can write the SVD as: $R=US V^\mathrm{T}$ or equivalently $R=\sum_{i=1}^{\min (n_1,n_2)}\sigma_i U_i \otimes V_i $  with $U_i$, $V_i$ being the $i^\mathrm{th}$ column vector of $U,V$. 
Similar to the convolution with $Q$ above, any convolution $(\cdot)*R$  can be computed by convolving each column of  $(\cdot)$ with $\sqrt{\sigma_i}U_i$, then convolving each row of that result with $\sqrt{\sigma_i}V_i$ and summing over $i=1,\dots,\min (n_1,n_2)$. Importantly, approximations can be made by truncating that sum after $m<\min (n_1,n_2)$.
We point out that the (singular value) decompositions of $Q$ and $R$ can be computed a priori since $Q$ and $R$ do not change during a simulation.

\begin{figure}[h!]
\centering
\includegraphics[width=\textwidth]{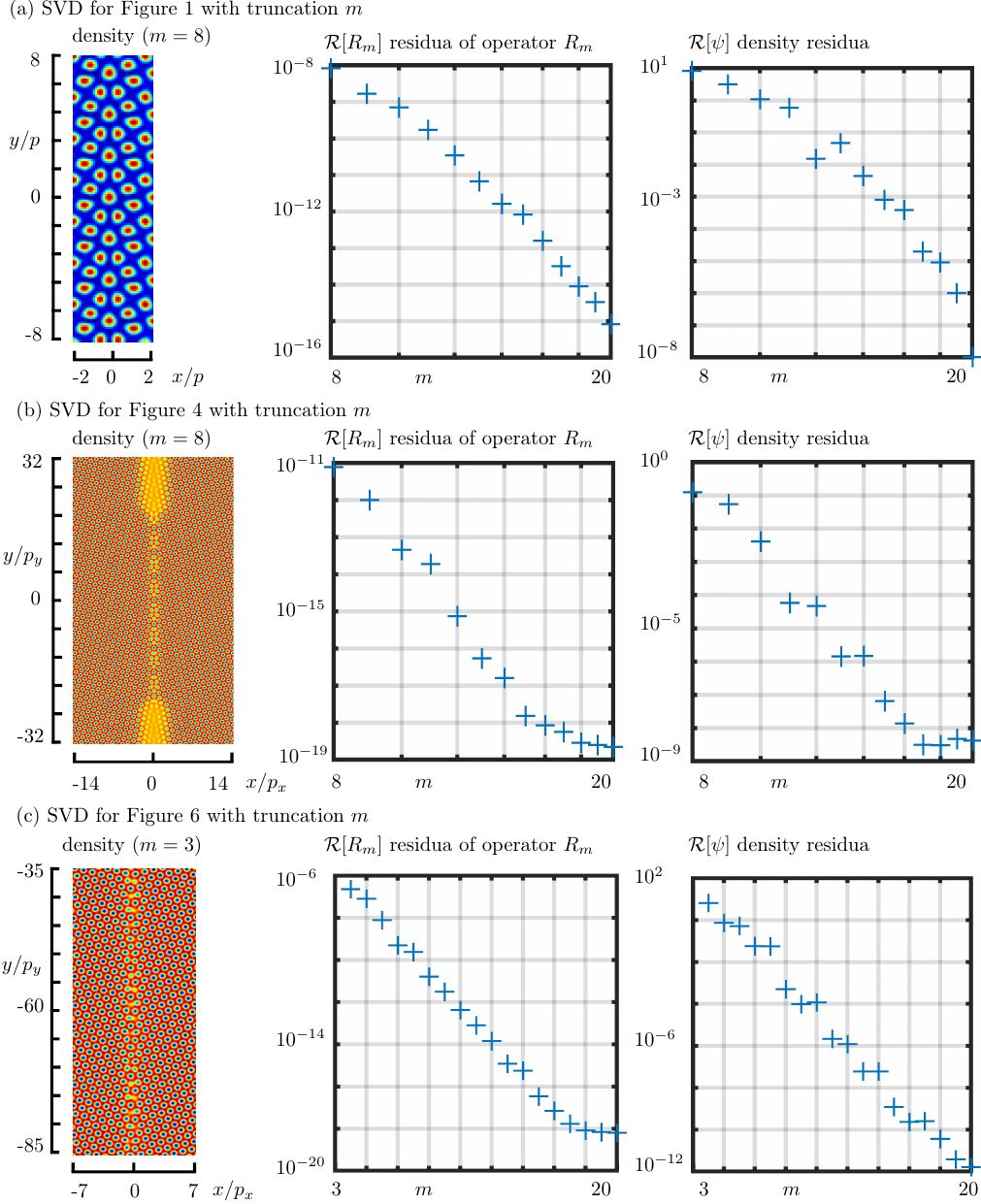}
\caption{SVD of the operator $R$ and effect of the truncation after $m$ singular values (see detail in the text): (a) Magnification of the relaxed structure as in Fig.~\ref{fig:GB}, evaluated for the truncation $m=8$ (left), numerical residua  $\mathcal{R}[R_m]$ for the truncation $R_m$, computed as the squared $L_2$ distance of $R_m$ to $R$ (middle) and numerical residua  $\mathcal{R}[\psi]$ computed as the squared $L_2$ distance of $\psi$ from the solution with SVD and the one obtained without SVD (right). Lengths are scaled with the atomic spacing along the $x$-axis and $y$-axis, $p_x$ and $p_y$.
(b) similar to (a) but for the setup illustrated in Fig.~\ref{fig:hybrid_scetch} (c) similar to (a) but for the setup illustrated in Fig.~\ref{fig:energy} ($\theta= 15^\circ$). 
	}
	\label{fig:svd}
    \end{figure}

Through numerical studies, we investigate how SVDs of $R$ influence the simulation results for the PFC model with \convalg~\eqref{eq:conv}. We consider the setup illustrated in Fig.~\ref{fig:GB} for the PFC model and in Figs.~\ref{fig:beats} and~\ref{fig:energy} for the hybrid-PFC model. For these setups, we compute the SVDs of $R$ and the respective residua for the truncations $R_m =\sum_{i=1}^{m< \min (n_1,n_2)}\sigma_i U_i \otimes V_i\approx R$ evaluated as the squared $L_2$ distance to $R$. We denote these residua by $\mathcal{R}[R_m]$.  Furthermore, we compute the residua of the density fields $\psi$ for a simulation with $R_m$ as the squared $L_2$ distance to the density field obtained by a simulation with $R$, denoted by $\mathcal{R}[\psi]$.
In Fig.~\ref{fig:svd} we plot the $\mathcal{R}[R_m]$ and $\mathcal{R}[\psi]$ with $m \leq 20$ for: (a) the setup as in Fig.~\ref{fig:GB} with fixed $\Omega_x/p\approx 64$; (b) the setup in Fig.~\ref{fig:hybrid_scetch} with fixed  $\omega=[-3.4,3.4]p_x\times [-23.9,20.7]p_y$, $\widetilde{\omega}=[-8.3,8.3]p_x\times [-28.6,25.5]p_y$; (c) the setup in Fig.~\ref{fig:energy} ($\theta= 15^\circ$) with fixed   $\omega=[-11.7,11.7]p_x\times [-85,85]p_y$, $\widetilde{\omega}=[-13.8,13.8]p_x\times [-85,85]p_y$ (only the central GB is modeled with local PFC accuracy within the simplified hybrid-PFC framework in this setting). We find a sharp decrease of all evaluated residua for increasing $m$. A comparison of Fig.~\ref{fig:svd}(a) ($R_8: 198$s run-time) with Fig.~\ref{fig:GB}(a) ($1548$s run-time, $\approx 8 \times$ slower), Fig.~\ref{fig:svd}(b) ($R_8: 20563$s run-time) with Fig.~\ref{fig:hybrid_scetch}(b) ($58568$s run-time, $\approx 3 \times$ slower) and Fig.~\ref{fig:svd}(c) ($R_3: 1675$s run-time) with Fig.~\ref{fig:energy}(b) ($7265$s run-time, $\approx 4 \times$ slower) shows that the truncations lead to simulation results in qualitative good agreement to the untruncated reference solutions while allowing for significantly faster computations.

\subsection{Refinement and domain studies}
\label{sec:refinement_domain}

\subsubsection{Refinement studies}

\begin{figure}[h!]
\centering
\includegraphics[width=\textwidth]{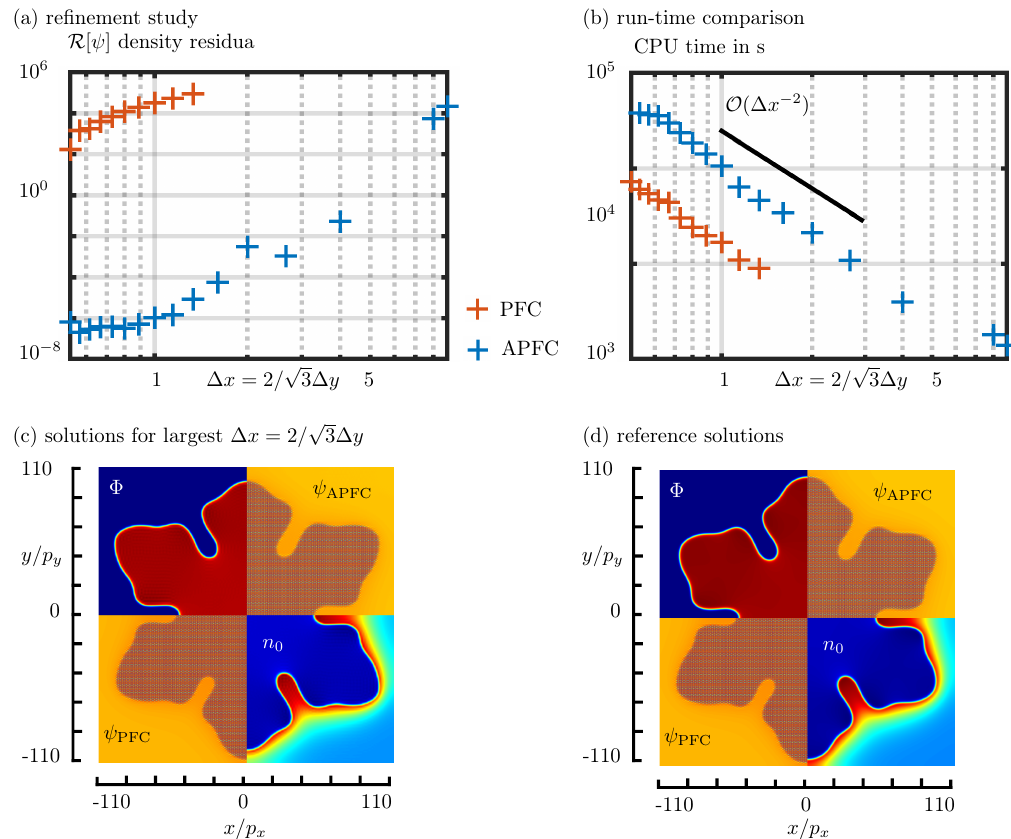}
\caption{Refinement and run time study for the APFC and PFC model. The setup illustrated in Fig.~\ref{fig:pfc_apfc} is considered. (a) numerical density residua $\mathcal{R}[\psi]$ for different grid sizes $\Delta x = 2/\sqrt{3}\Delta y$ (b) corresponding CPU times (c) solutions for largest grid sizes (d) reference solutions. For (c) and (d), the same (color)-scale as in Fig.~\ref{fig:pfc_apfc} is used.   Lengths are scaled with the atomic spacings along the $x$- and $y$-axis, $p_x$ and $p_y$.}
	\label{fig:refine}
    \end{figure}

In Fig.~\ref{fig:refine}(a) we present a numerical refinement study for the simulation setup illustrated in Fig.~\ref{fig:pfc_apfc} in the main text, to verify the important scaling difference between the APFC and the PFC. For both models we compute for different $\Delta x = 2/\sqrt{3} \Delta y$ the numerical residua $\mathcal{R}[\psi]$ as the squared $L_2$ distance of the density fields $\psi_\text{PFC}$ and $\psi_\text{APFC}$ (the latter via Eq.~\ref{eq:1mode}) to their respective refined solutions obtained by $\Delta x = 0.50$ and $\Delta y = 0.43$. To get comparable results, we interpolated the coarse solutions on a fine grid (as postprocessing) and obtained vanishing residua  $\mathcal{R}[\psi]$ for decreasing $\Delta x = 2/\sqrt{3} \Delta y$.  As shown in Fig.~\ref{fig:refine}(a) the PFC exhibits an exponential decay of $\mathcal{R}[\psi]$  for decreasing  $\Delta x = 2/\sqrt{3} \Delta y$ as expected by the Fourier pseudo-spectral method. For the APFC such an exponential decay is  visible when considering the residua of the amplitudes $\{\eta_m\}_{m=1}^3$ and local average density $\psi_0$ (not shown), which are the actual variables to solve for in the model, instead of the reconstructed $\psi_\text{APFC}$. For the PFC, a critical grid size of $\Delta x =1.33$, $\Delta y = 1.20$ is obtained (larger grids lead to numerical instabilities or significant differences with the reference solution). Due to the density field $\psi_\text{PFC}$, varying on the atomic length scale, the PFC is indeed limited by the required resolution to describe the crystal structure, namely to resolve density peaks properly. On the other hand, the APFC can be simulated up to $\Delta x = 8.90 $ and $\Delta y=7.70$; see Fig.~\ref{fig:refine}(a). The numerical grid spacing of the APFC is just limited by the width of the solid-liquid interface, which has to be resolved accordingly.
In Figs.~\ref{fig:refine}(c) and (d), we compare the solutions for the critical grid sizes to the reference solutions and obtain results in qualitative good agreement.  Comparing the run times of the APFC and the PFC model, one obtains an ideal quadratic scaling law (2D) of the CPU time with respect to $\Delta x$. For fixed $\Delta x = 1/\sqrt{3}\Delta y$, the PFC shows $\approx 7$ times faster performance than the APFC (for the PFC just $\psi_\text{PFC}$ needs to be updated, while for the APFC  $\{\eta_m\}_{m=1}^3$ and $\psi_0$ need to be updated). However, at respective grid size ensuring convergence (see above), we obtain  $\approx 10$ times faster performance of the APFC compared to the PFC (APFC: $\approx 1.4\cdot 10^3$ s CPU time, PFC: $\approx 1.3\cdot 10^4$ s CPU time).

\subsubsection{Domain study hybrid model}
\begin{figure}[h]
\centering
\includegraphics[width=\textwidth]{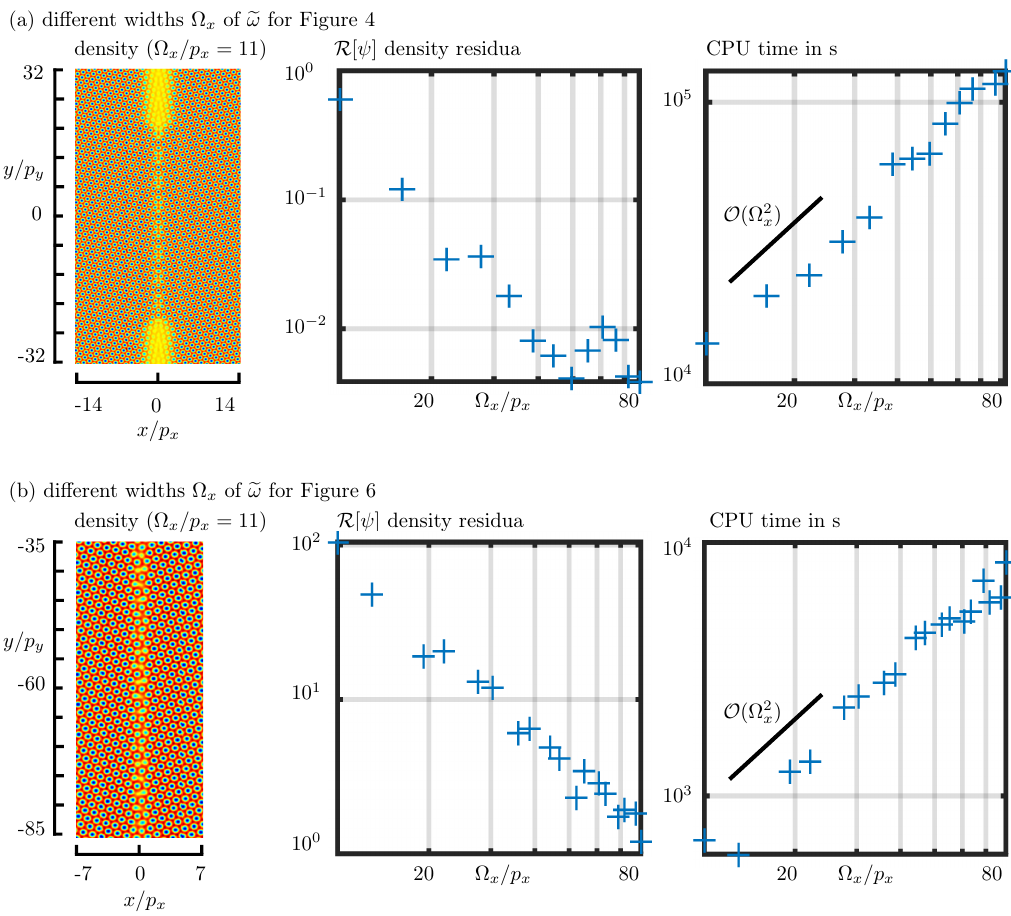}
\caption{Domain study for different widths $\Omega_x$ of $\widetilde{\omega}$ for the hybrid-PFC model: (a) setup as in Fig.~\ref{fig:hybrid_scetch} and (b) setup as in Fig.~\ref{fig:energy}. For both cases, the density profile for $\Omega_x=11p_x$ (left), density residua $\mathcal{R}[\psi]$ (middle), and required CPU time (right) are plotted. The same (color)-scales as in Fig.~\ref{fig:hybrid_scetch} and ~\ref{fig:energy} are used. Lengths are scaled with the atomic spacings along the $x$- and $y$-axis, $p_x$ and $p_y$.}
	\label{fig:omega}
    \end{figure}

For the setups illustrated in Fig.~\ref{fig:hybrid_scetch} and Fig.~\ref{fig:energy}, we vary the width of the domain $\widetilde{\omega}$ within the simplified hybrid-PFC framework while letting all other quantities unchanged. For  the setup illustrated in Fig.~\ref{fig:hybrid_scetch} we vary $\widetilde{\omega}=[-\Omega_{x}/2,\Omega_x/2]\times [-28.6,25.5]p_y$ within the range $11p_x\leq \Omega_x\leq 82 p_x$ for
fixed  $\omega=[-3.4,3.4]p_x\times [-23.9,20.7]p_y$ and the truncation $R_8$ according to the SVD presented in~\ref{sec:svd}. For the setup illustrated in Fig.~\ref{fig:energy} ($\theta= 15^\circ$) we vary $\widetilde{\omega}=[-\Omega_{x}/2,\Omega_x/2]\times [-85,85]p_y$ within the range $11p_x\leq \Omega_x\leq 82 p_x$ for
fixed  $\omega=[-0.7,0.7]p_x\times [-85,85]p_y$ and the truncation $R_3$; see ~\ref{sec:svd}.  
In Fig.~\ref{fig:omega}, we plot the numerical residua $\mathcal{R}[\psi]$ evaluated as  squared $L_2$ distance of the resulting density fields with respect to a reference solution (obtained by $\Omega_x=82p_x$) within the domain $\omega$. For increasing $\Omega_x$ vanishing $\mathcal{R}[\psi]$ and a quadratic increasing CPU time are observed. 

\subsection{Interpolation/Demodulation techniques}
\label{sec:interpol}
Having its origin in image processing, the so-called Fourier zero-padding method is a simple technique to numerically interpolate between (uniform) grids of different grid sizes in real space by a $\mathrm{sinc}$ interpolation, or equivalently, by convolutions with Dirichlet kernels~\cite{rabiner1975theory,fraser1989interpolation,smit1990efficient}. Suppose a given vector $A\in \mathbb{R}^n$ should be interpolated on a finer grid $m>n$. This can be achieved by adding  $m-n$ zeros to the discrete Fourier transform of $A$, denoted by $\widehat{A}=[\widehat{A}_1,\dots,\widehat{A}_n]\in \mathbb{C}^n$ so that  $\widehat{B}:=[\widehat{A}_1,\dots,\widehat{A}_{\lfloor n/2 \rfloor},0,\dots,0,\widehat{A}_{\lfloor n/2 \rfloor +1},\dots, \widehat{A}_n]\in \mathbb{C}^m$. The desired interpolated version of $A$ (denoted by $B\in \mathbb{R}^m$) can now be obtained by inversely Fourier transform $\widehat{B}$. By inverting the scheme above (starting from $B\in \mathbb{R}^m$ and removing the $m-n$ entries $\widehat{B}_{\lfloor n/2 \rfloor+1} \dots \widehat{B}_{\lfloor n/2 \rfloor +1+m-n}$ from $\widehat{B}$), one obtains the interpolation from a fine to a coarse grid. Throughout this manuscript, we used this technique globally on the whole computational grid for illustrating purposes, see Figs.~\ref{fig:pfc_apfc}, \ref{fig:hybrid_scetch}(a), \ref{fig:energy}(c), \ref{fig:refine}(c) as well as locally on $\widetilde{\omega}$ for our hybrid-PFC model; see Fig.~\ref{fig:hybrid_scetch}(c)-(e).

To guarantee a consistent coupling between the PFC and the APFC within the hybrid-PFC framework, one needs to translate the amplitudes $\{\eta_m\}_{m=1}^3$ as well as the local average density $\psi_0$ to the density $\psi$ and vice versa. While the (1-Mode) approximation of $\psi$ through  $\{\eta_m\}_{m=1}^3$ and $\psi_0$ is rather simple to compute, see Eq.~\eqref{eq:1mode}, the computation of $\{\eta_m\}_{m=1}^3$ and $\psi_0$, resulting from the PFC density needs a local demodulation. For small rotation angles, one can demodulate the density field via~\cite{skogvoll2021stress,skogvoll2022phase}
\begin{equation}
\label{eq:demodulation}
\begin{split}
    \eta_m &= \mathrm{e}^{-i  q_j^m r_j}\mathscr{F}^{-1}\left[\mathrm{e}^{-2\pi a_j^2 (k_j-q^n_j)^2 }\mathscr{F}\left[ \psi \right] \right]\quad \forall m=1,\dots, 3,\\
\psi_0 &= \mathscr{F}^{-1}\left[\mathrm{e}^{-2\pi a_j^2 k_j^2 }\mathscr{F}\left[ \psi \right] \right]
    \end{split}
    \end{equation}
with $\mathscr{F}$ the Fourier transform, $\mathscr{F}^{(-1)}$ the inverse Fourier transform, filter widths $a_1=p_x/2$, $a_y=p_y/2$ and implying Einstein summation convention. However, when large rotation angles $\theta$ are considered, as for the setups according to Fig.~\ref{fig:beats} and~\ref{fig:energy}, the demodulation~\eqref{eq:demodulation} needs to be done with respect to a rotated reference. This can be obtained by considering rotated reciprocal-lattice vectors~\cite{salvalaglio2017controlling,salvalaglio2020coarse}
\begin{equation*}
    \delta \boldsymbol{q}^m = \left[\begin{array}{cc}
         \cos \theta -1 & -\sin \theta \\
        \sin \theta &\cos \theta -1 
    \end{array} \right] \boldsymbol{q}^m\quad \forall m=1,\dots,3
\end{equation*} 
and use them for the demodulation
\begin{equation*}
        \eta_m= \mathrm{e}^{-i  q_j^m r_j}\mathscr{F}^{-1}\left[\mathrm{e}^{-2\pi a_j^2 (k_j-\delta q^m_j)^2 }\mathscr{F}\left[ \psi \right] \right] \forall m=1,\dots, 3
\end{equation*}
(the computation of $\psi_0$ remains unchanged).
For the setups shown in Fig.~\ref{fig:beats}  and Fig.~\ref{fig:energy}, one needs to account for two rotation angles $\pm \theta$, acquiring separate demodulations within the domain $\widetilde{\omega}$ for $+\theta$ and $-\theta$, respectively. Strategies can also be devised to adapt the angle for the demodulation during the simulation. Note that the hybrid-PFC model thus allows for considering different orientations of reference lattices for problem subdomains. This is prevented for most of the APFC models as one should then define how the reciprocal-space vectors rotate at the interface between these grains, still correctly describing the microscopic density and the resulting GB structure (a problem addressed only partially in Refs.~\cite{bervcivc2018adaptive,bervcivc2020enabling}). For the hybrid-PFC model, such a region is where the accurate PFC model is considered without the need for ad-hoc extension of the APFC model (see also Fig.~\ref{fig:energy}).

\end{document}